\documentclass[preprints,article,accept,moreauthors,pdftex]{Definitions/mdpi}

\firstpage{1} 
\pubvolume{1}
\issuenum{1}
\articlenumber{0}
\pubyear{2021}
\copyrightyear{2020}
\datereceived{} 
\dateaccepted{} 
\datepublished{} 
\hreflink{https://doi.org/} 

\Title{Role of magnetic fields in ram pressure stripped galaxies}

\TitleCitation{Magnetic fields in ram-pressure stripped galaxies}

\Author{Ancla M\"{u}ller $^{1}$\orcidA{}, Alessandro Ignesti $^{2}$\orcidB{}, Bianca Poggianti $^{2}$\orcidC{}, Alessia Moretti $^{2}$\orcidD{},\\ Mpati Ramatsoku $^{3,4}$\orcidF{}, and Ralf-Jürgen Dettmar $^{1}$\orcidE{} }

\AuthorNames{Firstname Lastname, Firstname Lastname and Firstname Lastname}

\AuthorCitation{Müller et al.}

\address{%
$^{1}$ \quad Ruhr University Bochum, Faculty of Physics and Astronomy, Astronomical Institute, Universit\"atsstr.150, 44801 Bochum, Germany\\
$^{2}$ \quad INAF-Osservatorio Astronomico di Padova, Vicolo dell'Osservatorio 5, I35122 Padova, Italy\\
$^{3}$ \quad Department of Physics \& Electronics, Rhodes University, Makhanda, 6139, South Africa\\
$^{4}$ \quad INAF-Osservatorio Astronomica di Cagliari, Via della Scienza 5, I-09047 Selargius (CA), Italy}

\corres{Correspondence: amueller@astro.rub.de}

\abstract{Ram-pressure stripping is a crucial evolutionary driver for cluster galaxies and jellyfish galaxies, characterized by very extended tails of stripped gas, are the most striking examples of it in action. Recently, those extended tails are found to show on-going star formation raising the question how the stripped, cold gas can survive long enough to form new stars outside the stellar disk. In this work, we summarize the most recent results achieved within the GASP collaboration to provide a holistic explanation for this phenomenon. We focus on two textbook examples of jellyfish galaxies, JO206 and JW100 for which, via multi-wavelength observations from radio to X-ray and numerical simulations, we have explored the different gas phases (neutral, molecular, diffuse-ionized, and hot). Based on additional multi-phase gas studies we now propose a scenario of stripped tail evolution including all phases that is driven by a magnetic draping sheath, where the intracluster turbulent magnetized plasma condenses onto the galaxy disk and tail and produces a magnetized interface that protects the stripped galaxy tail gas from evaporation. In such a scenario, the accreted environmental plasma can cool down and eventually join the tail gas, hence providing additional gas to form stars.  The implications of our findings can shed light on the more general scenario of draping, condensation, and cooling of hot gas surrounding cold clouds that is fundamental in many astrophysical phenomena.}

\keyword{Galaxies: magnetic fields; Galaxies: evolution; Galaxies: ISM; Radio continuum: galaxies; Polarization} 

\begin{document}

\section{Introduction}

On cosmic scales, the star formation activity in galaxies has strongly declined since z $\sim 2$ such that a large number of galaxies have evolved into passive galaxies while the star formation rate at fixed stellar mass has decreased \citep{Noeske2007}.
To understand the drop in the star formation rate density of the Universe \citep{Madau2014} it is necessary to study the processes of gas acquisition and loss, which can be divided into internal and external mechanisms. Gas consumption as well as stellar or Active Galactic Nuclei (AGN) feedback processes feeding the galaxy halo can lead to a decrease in the star formation rate.
There are several external mechanisms depleting the gas content of a galaxy \citep{Boselli2006}, and their efficiency depends on whether the galaxy is isolated, or is part of a group, or a cluster.
One of the most efficient environmental mechanisms is ram pressure stripping that has been observed via neutral hydrogen studies in galaxy clusters \citep{Cayatte,Kenney2004,Chung09,Jaffe2015,Scott18}. A varied situation is found in galaxy groups \citep{William1987,Montenegro2001,Serra2013}.

More recently, extreme ram pressure stripped galaxies with long gaseous tails, named jellyfish galaxies, were found moving at high speed in a variety of galaxy clusters \citep[e.g.,][]{Bellhouse2017,Boselli21}.
Significant on-going star formation has been detected within galactic disks and far out into the tails \citep{Cortese07,Poggianti2019b}. 
This demonstrates that the process of gas removal is not always associated with the immediate quenching of star formation. On the contrary, the exact interplay between gas stripping and star formation activity might vary from galaxy to galaxy depending on their size, location and orbit within a cluster, as well as on the cluster properties. Understanding this interplay is crucial to fully understand how star formation is eventually quenched in dense environments.

Among the most compelling open questions are the origin of the in-situ star-formation in ram pressure stripped tails, and whether magnetic fields promote star-formation activity in the tails.
Magnetic fields can play a key role in star formation processes: on parsec scales, molecular clouds collapse through magnetic field transport of angular momentum \citep{Crutcher2012}. Spiral galaxies show large-scale magnetic fields along and between the optical spiral arms and in their halos \citep{Heesen11,Beck2013}, anchoring the magnetic field to molecular clouds \citep{Li2011}. 
Until recently the magnetic field of ram pressure stripped galaxies was only observed in the close vicinity of galaxy discs in nearby galaxy clusters \citep[e.g.,][]{Gavazzi1995,Vollmer2010,Chen}. By measuring the radio continuum brightness and spectral index trend in direction of the galaxy tail these works found an excess in radio emission that cannot be explained by the current star-formation in the H$\alpha$ emitting knots. 
This can be the result of the ram-pressure stripping process, in agreement with the steepening of the spectral index with distance to the galaxy disks, which is expected in a scenario where electrons travel from the disk to the tail and release their energy, indicating standard electron cooling. Based on such standard cooling assumptions, \citep{Gavazzi1995} predicted a magnetic field strength of about $5\,\mu$G in the galaxy tails assuming several parameters. This value is found to agree with more recent extreme ram pressure stripped tails \citep{Mueller2021,Ignesti21}. 
\citet{Mueller2021} studied for the first time the magnetic field strength and configuration of both the disk and the tail of an extreme ram pressure stripped galaxy, JO206. 

Long co-spatial radio continuum and H$\alpha$ tails in several more ram pressure stripped galaxies have been found \citep[][M\"uller et al., in prep]{Gavazzi1995,Poggianti2019a,Chen,Ignesti21}, suggesting that such tails are common.
The number of known spiral galaxies with radio continuum tails, i.e., evidence of a magnetic field extended beyond the stellar disk for tens of kpc, has greatly increased recently thanks to the progress of the LOFAR Two-metre Sky Survey \citep[LoTSS,][]{Shimwell_2017}. Based on the LoTSS observations, \citep{Roberts2021,Roberts_2021b} reported the discovery of hundreds of radio jellyfish 
in low-$z$ clusters and groups. These results indicate that the formation of radio tails is ubiquitous in clusters, as well as showing that LOFAR can be a powerful instrument for identifying ram pressure stripped galaxies across extremely wide fields.
 
\section{The GASP Survey}

The GAs Stripping Phenomena in galaxies (GASP\footnote{\url{https://web.oapd.inaf.it/gasp/}}) survey \citep{Poggianti2017} has assembled a multi-wavelength data set of jellyfish galaxies in low-z clusters. Within a Multi Unit Spectroscopic Explorer (MUSE) European Southern Observatory (ESO) large program GASP has obtained integral-field optical spectroscopy of 114 galaxies (ram-pressure stripped galaxy candidates and control sample) at $z=0.04-0.07$ with masses from $10^9$\,M$_{\odot}$ to $10^{11.5}$\,M$_{\odot}$ in clusters, groups, filaments, and isolated. These sources were selected based on unilateral debris or tails in B-band images (and no signs of it for the control sample) detected in the WINGS and OmegaWINGS survey \citep{Poggianti2016}. 

To study the cold neutral gas in the galaxy disk and tails GASP also assembled H\textsc{i} observations from the Karl G. Jansky Very Large Array (VLA), MeerKAT, and Australia Telescope Compact Array (ATCA) of a sub-sample of GASP galaxies \citep[][Tirna et al., submitted]{Ramatsoku2019,Ramatsoku2020,Deb2020}. We further observed the radio continuum using the same instruments and also the LOw Frequency ARray (LOFAR) \citep[][Ignesti et al. submitted, M\"uller et al., in prep]{Poggianti2019a,Mueller2021}. To study the magnetic field structure we performed deep polarization observations with the VLA and ATCA for five ram-pressure stripped galaxies, however, only one observation is already completed \citep[JO206][see below]{Mueller2021}. 
GASP also collected Atacama Pathfinder EXperiment (APEX) CO(2-1) \citep{Moretti2018}, Atacama Large Millimeter/submillimeter Array (ALMA) CO(2-1) and CO(1-0) \citep{Moretti2020} data to study the molecular gas component in the disks and tails, Ultra Violet (UV) data with Ultra-Violet Imaging Telescope (UVIT) on the AstroSat, and UV to I-band data with the Hubble Space Telescope (HST).
Finally, GASP has been awarded a deep Chandra X-ray Observatory (CXO) observation of JO206 and utilised archival CXO data to study the cluster environment \citep{Mueller2021} and X-ray emission of extreme ram-pressure stripped galaxies \citep{Poggianti2019a, Campitiello2021}.\footnote{For any more details and the collection of publications see \url{https://web.oapd.inaf.it/gasp/}}

By assembling the different gas phases and magnetic field properties of a sub-sample of GASP galaxies we aim to study galaxy evolution addressing the following questions:
What is the baryonic cycle between the different gas phases and the star formation in the disk and tails of ram pressure stripped galaxies? Where, how, and why is gas removed from galaxies and what is the effect on the galaxy star-formation?
Is there a possible connection between ram pressure and the appearance of AGNs in the galaxy centers? Does ram pressure cause the quenching of star-formation on long term scales and the transformation to passive early type galaxies? What is the impact of ram-pressure in groups and filaments?
In this work, we will focus on the recent results from the GASP collaboration about the properties of the different gas phases and how they can survive long enough in the tails to evolve into stars focused on the recent multi-phase gas, radio continuum and linear polarization results collected within GASP.
\begin{figure*}[t]
\centering
\includegraphics[scale=0.56]{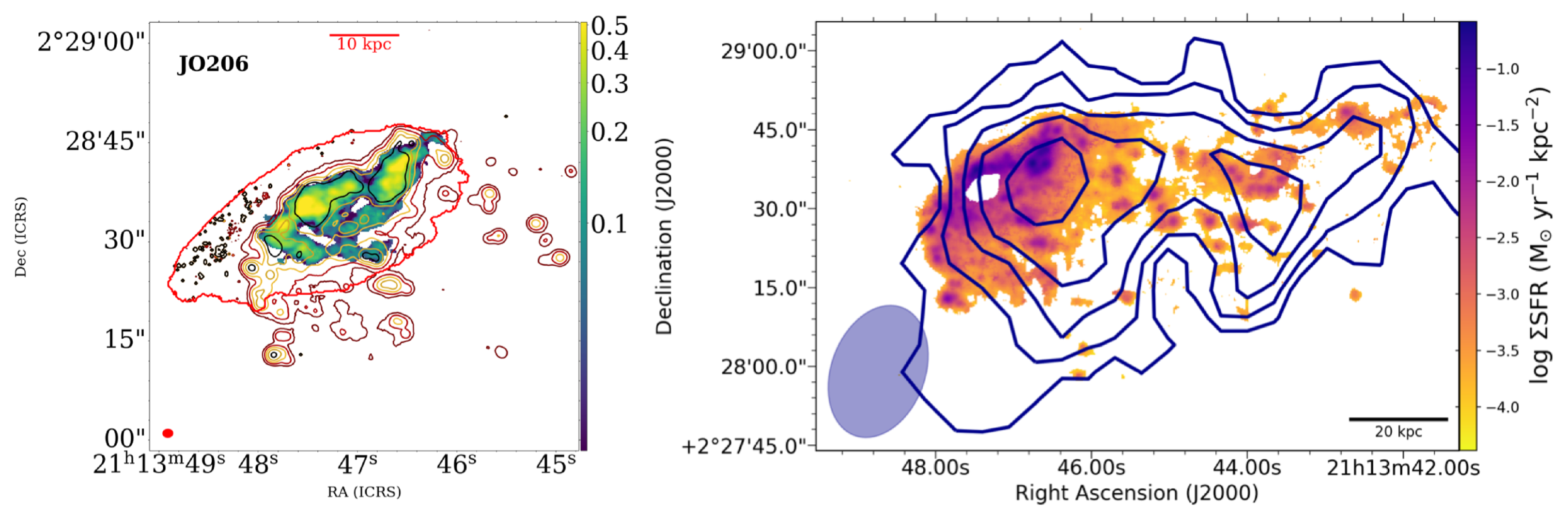}
\caption{Multi-phase gas components of JO206: \textbf{Left:} Molecular gas CO(1-0) image in units Jy/beam.km/s and H$\alpha$ ionised gas contours taken from \citet{Moretti2020}. The stellar disk is shown as a red contour. Reproduced with permission from A. Moretti, published by AAS (ApJ), 2020. \textbf{Right}: Star formation rate surface density image and H\textsc{i} neutral gas contours taken from \citet{Ramatsoku2019}. The H\textsc{i} beam is shown in the left corner. \label{JO206_multi}}
\end{figure*}
\section{Multi-phase gas}

Jellyfish galaxies are found to be H\textsc{i} deficient \citep[][Deb et al., in prep; respectively]{Ramatsoku2019, Ramatsoku2020, Deb2020} but contain star-forming clumps with reservoirs of H$_2$ $\sim$ 4 times higher than expected for normal star-forming galaxies of similar mass \citep[][]{Moretti2018,Moretti2020}. In addition, the amount of neutral + molecular gas is found to be comparable to normal star-forming galaxies of similar galaxy mass indicating 1) an effective conversion from H\textsc{i} to H$_2$ \citep{Moretti2020} and 2) either no significant evaporation of the cold gas into the ICM, or an effective accretion and mixing of hot ICM that can cool providing an additional reservoir of gas, or (probably) a combination of both supported by the magnetic field configuration in the galaxy tails (see below).

The diffuse ionized gas has been studied for 71 GASP galaxies, both ram-pressure stripping candidates and a control sample \citep{Tomicic}. We find that in the stripped tails (and even more significantly in the most extended tails such as JO206 and JW100), the OI/H$\alpha$ ratio reaches values that cannot be explained by the current star-formation alone. It is likely that at least a part of the ionization is caused either by shocks (no direct evidence found in the X-ray observations yet) or accretion of hot ICM gas and cooling and/or mixing.
\begin{figure*}[t]
\centering
\includegraphics[scale=0.54]{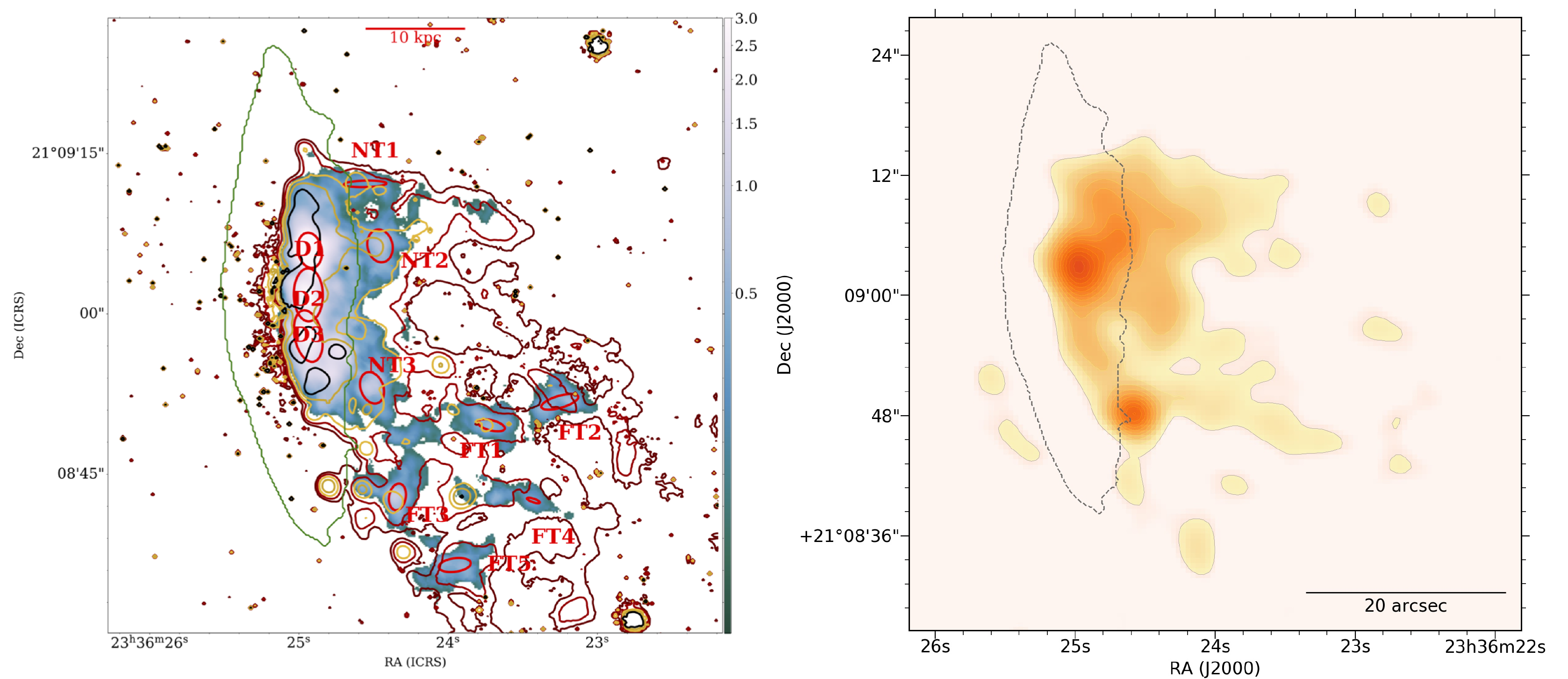}
\caption{Multi-phase gas component of JW100: \textbf{Left:} Molecular gas CO(2-1) image in units Jy/beam.km/s and H$\alpha$ ionized gas contours taken from \citet{Moretti2019}. The stellar disk is shown in green contours. Reproduced with permission from A. Moretti, published by AAS (ApJ), 2019. \textbf{Right:} Hot gas (X-ray) in orange and stellar disk is shown in black contours.\label{JW100_multi}}
\end{figure*}
Studies of GASP galaxies \citep[][]{Poggianti2019a,Campitiello2021}, as well as other objects \citep[e.g.,][]{Sun2021} revealed that jellyfish galaxies can host a hot phase ($T\simeq10^6-10^7$ K) with a metallicity tentatively consistent with the ICM values ($Z\simeq0.3-0.4$ $Z_\odot$) that emits X-rays via thermal bremsstrahlung and closely follows the morphology of the stripped H$\alpha$-emitting tail. These results are in agreement with those reported for other jellyfish galaxies \citep[see][and references therein]{Sun2021}.
The high temperature of this phase and the generally low metallicity might suggest that this X-ray emitting phase is mostly composed by cooled ICM, thus supporting the idea that jellyfish galaxies are actually able to accrete hot, magnetized ICM during their orbits of the cluster. However, the nature of the interplay between ISM and ICM which ultimately leads to the formation of the X-ray extended emission is still unclear.

On a more general view, the most remarkable jellyfish galaxies, that experience in-situ star formation in extended gas tails, are found to be the most massive disk galaxies in the clusters and move at very high velocities with respect to the intracluster medium \citep[][]{Jaffe2018}. The high total mass, therefore, allows the galaxy to hold onto its gas, until it has reached the densest part of the ICM, which is also an important factor to understand the lifetime and gas phases in the extended ram-pressure stripped tails. The high ICM pressure can effectively build a magnetic sheath protecting the gas from evaporation into the ICM at the tail edges (observed in case of JO206, see below). In the followings we summarize the most relevant results achieved for two of the most studied galaxies of the gasp survey, JO206 and JW100.

JO206 is a textbook case galaxy to study ram-pressure stripping due to its 90\,kpc-long H$\alpha$-emitting ionized gas tail \citep{Poggianti2017}. JO206 is part of the low-mass cluster IIZW108 ($\sigma =575\,$km/s) at a redshift of $z=0.05133$, it has a total stellar mass of $M_\star = 7.8 \times 10^{10}$\,M$_\odot$ and a star-formation rate of 4.8\,M$_\odot$/yr and 0.5\,M$_\odot$/yr in its disk and tail, respectively. It is falling into the cluster with a velocity of at least 1200\,km/s at a projected distance of 300\,kpc from the cluster center. The multiple star-forming clumps in the galaxy disk and tail were identified based on their H$\alpha$ emission. These clumps are generally found to coincide also with a reservoir of molecular gas \citep{Moretti2018,Moretti2020} (left panel in Fig. \ref{JO206_multi}) and neutral hydrogen \citep{Ramatsoku2019} (right panel in Fig. \ref{JO206_multi}). The molecular gas follows the clumpy structure of the star-forming knots. The lower spatial resolution H\textsc{i} observations show the presence of H\textsc{i} along the whole H$\alpha$-emitting tail. 
However, the total amount of neutral + molecular gas is found to be normal \citep{Moretti2020}. In \citet{Tomicic} we found an additional ionized gas component that cannot be associated with the current star-formation in the galaxy disk and tail.
\begin{figure*}[t]
\centering
\includegraphics[scale=0.70]{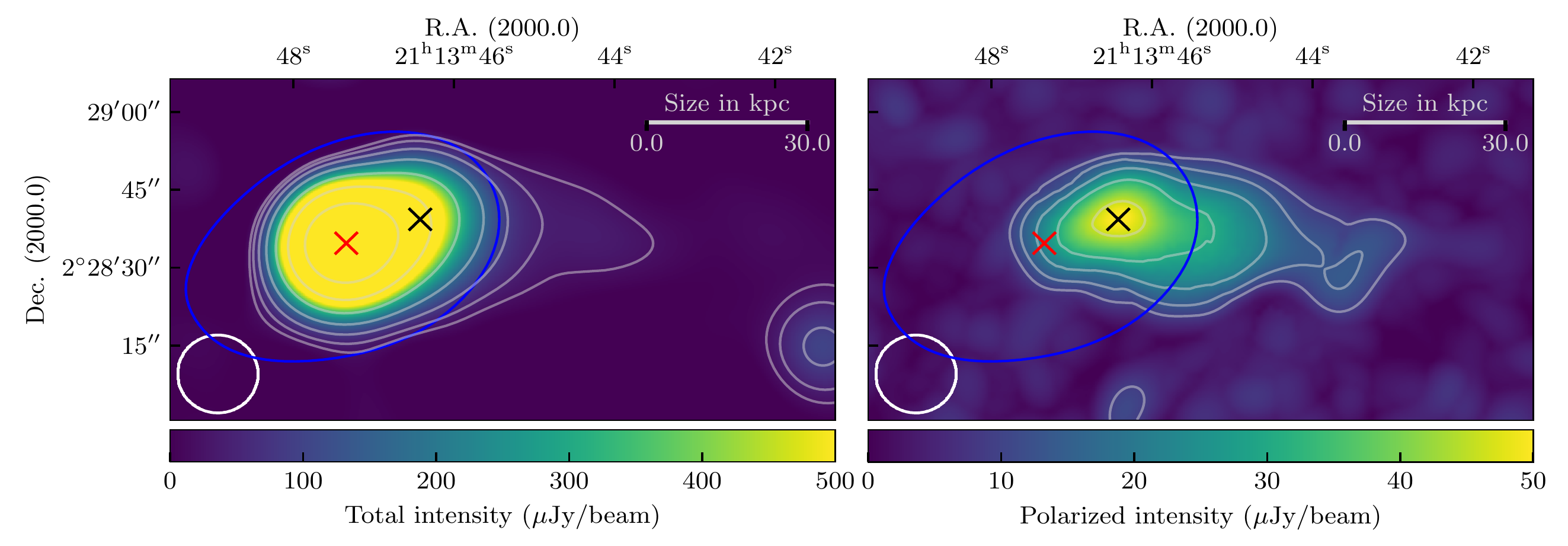}
\caption{Radio emission of JO206 at 2.7\,GHz adapted from \citet{Mueller2021}: \textbf{Left:} Radio continuum emission and radio continuum contours with its peak shown in red. The radio beam is shown in the left corner and the stellar disk convolved to the radio resolution is shown in blue contours. \textbf{Right:} Linear polarized emission and linear polarized emission contours with its peak shown in black. \label{JO206_radioemis}}
\end{figure*}
%

The jellyfish galaxy JW100, at a redshift of $0.06189$, is part of the A2626 cluster ($\sigma =650\,$km/s) \citep{Poggianti2019b}.
It is located at a projected distance of 100\,kpc from the cluster center and is falling into the cluster with a line-of-sight velocity of 1800\,km/s. The galaxy has a stellar mass of $M_\star = 2.9 \times 10^{11}$\,M$_\odot$, which makes it the most massive galaxy in the GASP sample, and forms effectively stars in its disk and tail, mostly in the southern tail region, with a star-formation rate of 3.2\,M$_\odot$/yr and 0.8\,M$_\odot$/yr, respectively. 
Multiple gas phases of JW100 are already coherently discussed in \citet{Poggianti2019a} for the galaxy disk and the 50\,kpc-long H$\alpha$-emitting gas tail. The star-forming regions coincide with H$\alpha$, UV, and CO clumps (left panel in Fig. \ref{JW100_multi}). However, not all observables are detected in all clumps at the same time and this can be interpreted as a star-formation sequence. The amount of molecular gas is also found to be quite high \citep[][]{Moretti2020} and the diffuse ionized gas can not be explained by the star-formation alone \citep{Tomicic}. 
Finally, {\it Chandra} X-ray observations revealed that JW100 has an extended X-ray emission elongated along the tail (right panel in Fig. \ref{JW100_multi}). The temperature of this component is $\sim$1 keV, which is not consistent with neither the temperature of the surrounding ICM \citep[$kT=3.5$ keV,][]{Ignesti_2018} or the typical hot ISM phase \citep[$kT\simeq0.1-0.2$ keV, e.g.,][]{Strickland_2004a}. Interestingly, the X-ray luminosity is higher than expected based on the current SFR, thus suggesting that the hot gas enveloping JW100 was not only heated up by the star formation.

\section{Radio continuum and polarization observations}
\label{sec:cont}

Ram pressure stripped galaxies have been observed in radio continuum over the past decades. Within GASP we conducted two case studies to grasp the role of radio continuum and the corresponding magnetic field strength not only in the host galaxy disks but also in the extended jellyfish tails of JO206 \citep{Mueller2021} and JW100 \citep{Poggianti2019a,Ignesti21}. In the first mentioned study, we were additionally able to study for the first time the magnetic field configuration far into the extended galaxy tail. Here, we provide a summary of both case studies showing only the most important observational findings. Theoretical implications, we drew from them are given in the next section.

%
\begin{figure*}[t]
\centering
\includegraphics[scale=0.95]{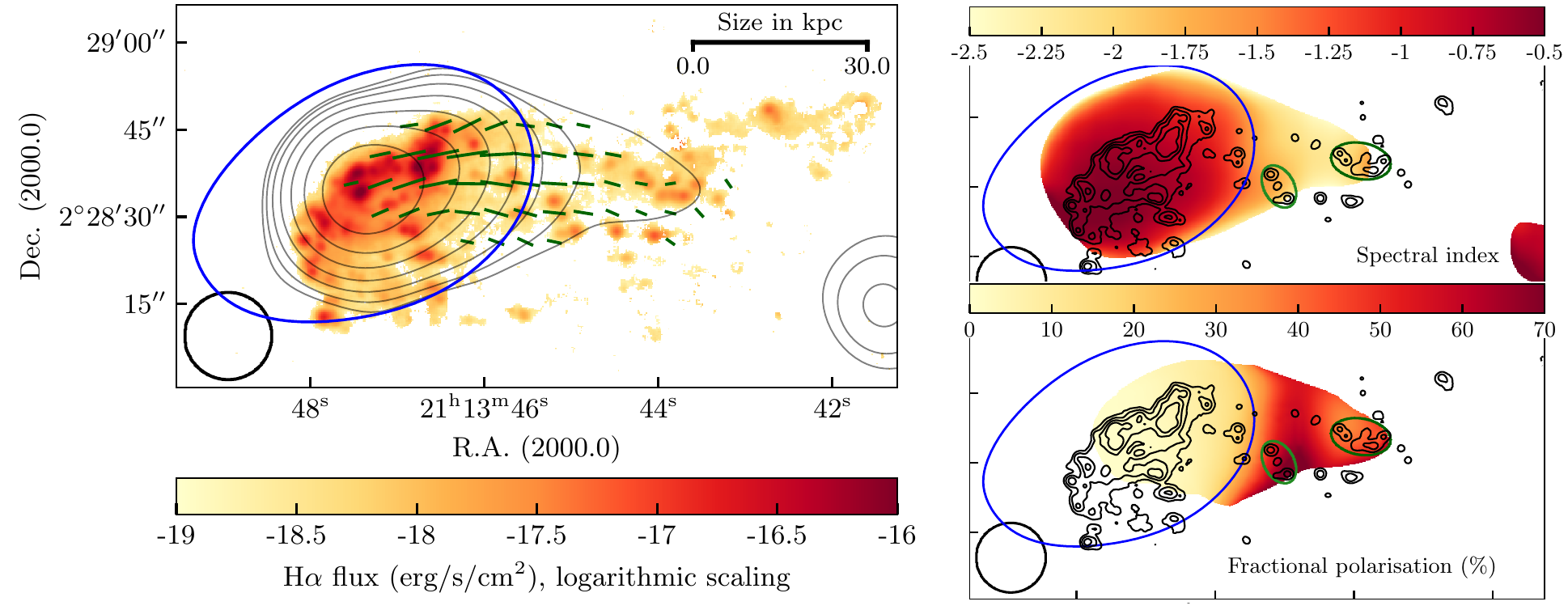}
\caption{Radio observables of JO206 at 2.7\,GHz adapted from \citet{Mueller2021}: \textbf{Left:} H$\alpha$ ionised gas image and magnetic field vectors in green. The radio beam is shown in the left corner and the stellar disk convolved to the radio resolution is shown in blue contours. \textbf{Top right:} Spectral index image between 2.7\,GHz and 1.4\,GHz and H$alpha$ ionized gas contours. \textbf{Bottom right:} Fractional polarization and H$\alpha$ ionized gas contours.
\label{JO206_radio}}
\end{figure*}
In \citet{Mueller2021} deep VLA C-array 2.7\,GHz in full polarization mode were combined with VLA C-array 1.4\,GHz observation [22] to study the spectral index and magnetic field properties in the 90-kpc H$\alpha$ emitting, ionized gas tail of JO206 selected from GASP.
In Figure \ref{JO206_radioemis} we show that the extent of the tail in radio continuum and polarization exceeds 40\,kpc at 2.7\,GHz, which is one of the most extreme ram-pressure stripped galaxy tail forming in-situ stars in the radio range reported in so far.
We find the radio disk properties to be in general agreement with star-forming galaxies -- a spectral index of $\alpha \approx -0.7$, a magnetic field strength of $B\approx 7\,\mu$G, and a polarized fraction of 2\,\% --, which is in agreement with previous studies, while the tail appears significantly different. The tail spectral index is found to be steepening with distance from the galaxy disk, reaching a mean value of about -2 (top right panel in Fig. \ref{JO206_radio}), indicating a significant aging of electrons through the tail. We investigated different methods (equipartition, electron cooling length, and magnetic draping) to restrict the tail magnetic field to $B\approx(2-4)\,\mu$G. The combination of high degree of fractional polarization of about 54\,\% (bottom right panel in Fig. \ref{JO206_radio}) and the derived polarization angle (corrected for Faraday rotation) revealed an aligned large scale magnetic field along the length of the tail in the stripping direction (left panel in Fig. \ref{JO206_radio}).
We identified star-forming knots in the tail (green regions in Fig. \ref{JO206_radio}) with a spectral index flattening and reduced fractional polarization that however are not able to disrupt the highly ordered magnetic field. The latter is aligned with the tail and illuminated by aged electrons with a steep spectral index. Such ordered magnetic fields may prevent heat and momentum exchange, and favor in-situ star-formation in the tail. In this scenario, star-formation in the tail should be a self-regulating process: a strong and widespread star-formation activity would disrupt the ordered magnetic field, which in turn is fundamental to allow the star-formation process to continue.
The study of JO206 is unique and of general interest to understand the impact of magnetic fields during ram pressure stripping and star-formation. So far, such a configuration of spectral index steepening, high fractional polarization, and aligned magnetic field was only found in head-tail galaxies (AGN driven jets experiencing ram pressure stripping) \citep[see][and references therein]{Mueller21}. This recent study of JO206 puts forward the idea that such a combination can also be relevant to understand the star-forming, ram-pressure stripped tails in galaxy clusters.
\citet{Mueller2021} also provide a comparison to magneto-hydrodynamical simulations to evaluate the origin of the radio emission. We provide this picture in the next section.

%
\begin{figure*}[t]
\centering
\includegraphics[scale=0.15]{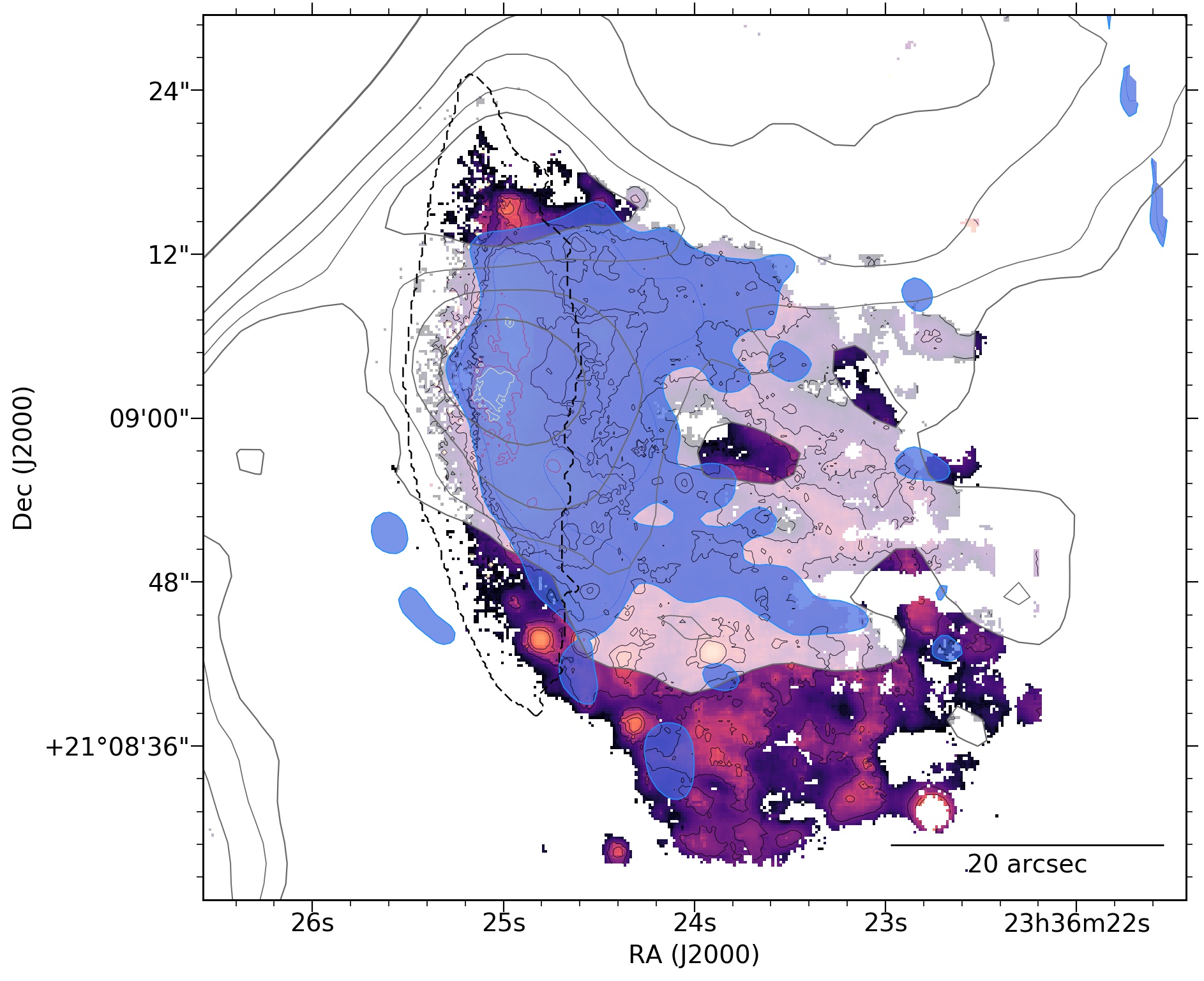}
\caption{Composite image of MUSE H$\alpha$ (red, 0.2$^{\prime\prime}$ resolution), Chandra X-ray (blue contour, 1.5$^{\prime\prime}$ resolution) and LOFAR radio continuum at 144\,MHz (grey contours at $3,6,12,24\times\textrm{rms}$, $\textrm{rms}=94\,\mu$Jy/beam, $8.8^{\prime\prime}\times6.4^{\prime\prime}$ resolution) of JW100 from \citet{Ignesti21}. Reproduced with permission from A. Ignesti, published by AAS (ApJ), 2021. \label{JW100_radio}}
\end{figure*}
JW100 also shows an extended, non thermal radio emission. Although a detailed polarimetric analysis has not been carried out yet, its emission has been detected in a wider range of frequencies than JO206, extending from 144 MHz to 5.5 GHz \citep[][]{Gitti_2013,Ignesti_2017,Ignesti_2020}. A combined analysis has recently been investigated within the GASP collaboration involving, for the first time, LOFAR, MeerKAT and VLA observations \citep{Ignesti21}. Overall, the radio emission follows the morphology of both the H$\alpha$ and X-ray emission (Fig. \ref{JW100_radio}). It is truncated with respect to the stellar disk, and extends linearly along the stripped tail, thus, supporting the idea that the non-thermal radio emission (due to the presence of the magnetic field) is affected by the ram pressure stripping.  \citep{Ignesti21}. The synchrotron spectrum revealed that the relativistic electrons are accelerated within the disk and then displaced by the ram pressure stripping, in according with the ram pressure stripping scenario. Thus, similarly to the case of JO206, a spectral steepening trend has also been observed for JW100. Moreover, the comparison between the continuum spectrum and the H$\alpha$ emission allowed us to constrain the time-scale of the star formation quenching \citep{Ignesti21}.

In agreement with the findings for JW100 and JO206 the radio continuum in ram pressure stripped galaxies is found to be co-located with H$\alpha$ not only within the disk but also with the extended, H$\alpha$ emitting tails, namely JO204, JO194, JO175, and JO147 (M\"uller et al., in prep).

\section{Theoretical view}

Theoretical assumptions, simulations, and their comparison to observations of ram-pressure stripped galaxies have provided several approaches to explain the enhanced radio continuum in the galaxy disks and tails, the appearance of the different gaseous phases, and the magnetic field strength and configuration.
In the literature, there have been a few simulations of ram-pressure stripped galaxies including magnetic fields over the last decade. \citet{Tonnesen} find the stripping rate of the disk gas not to be significantly affected by the galaxy magnetic field strength and morphology while \citet{Martinez} propose a reduced compressibility of the magnetized plasma that makes the tail gas less clumpy and filamentary-like in comparison to non-magnetized studies.
\citet{Vollmer07} find asymmetric polarized ridges that can be explained through adiabatic compression of galaxy magnetic fields at the leading edge. In addition, galaxies are expected to sweep up the intracluster magnetic field as they orbit inside a galaxy cluster and develop a magnetic draping sheath \citep{Dursi}. Such a sheath is lit up with cosmic rays originating from supernovae and generate coherent polarized emission at the leading edges of the galaxy \citep{Pfrommer10}. Importantly, \citet{Ruszkowski} propose that simulated galaxies interacting with a magnetized ICM show highly ordered magnetic fields in the tails that suppress strong Kelvin-Helmholz instabilities at the interface between the stripped gas and the ICM \citep{Berlok}. 
\begin{figure*}[t]
\centering
\includegraphics[scale=0.555]{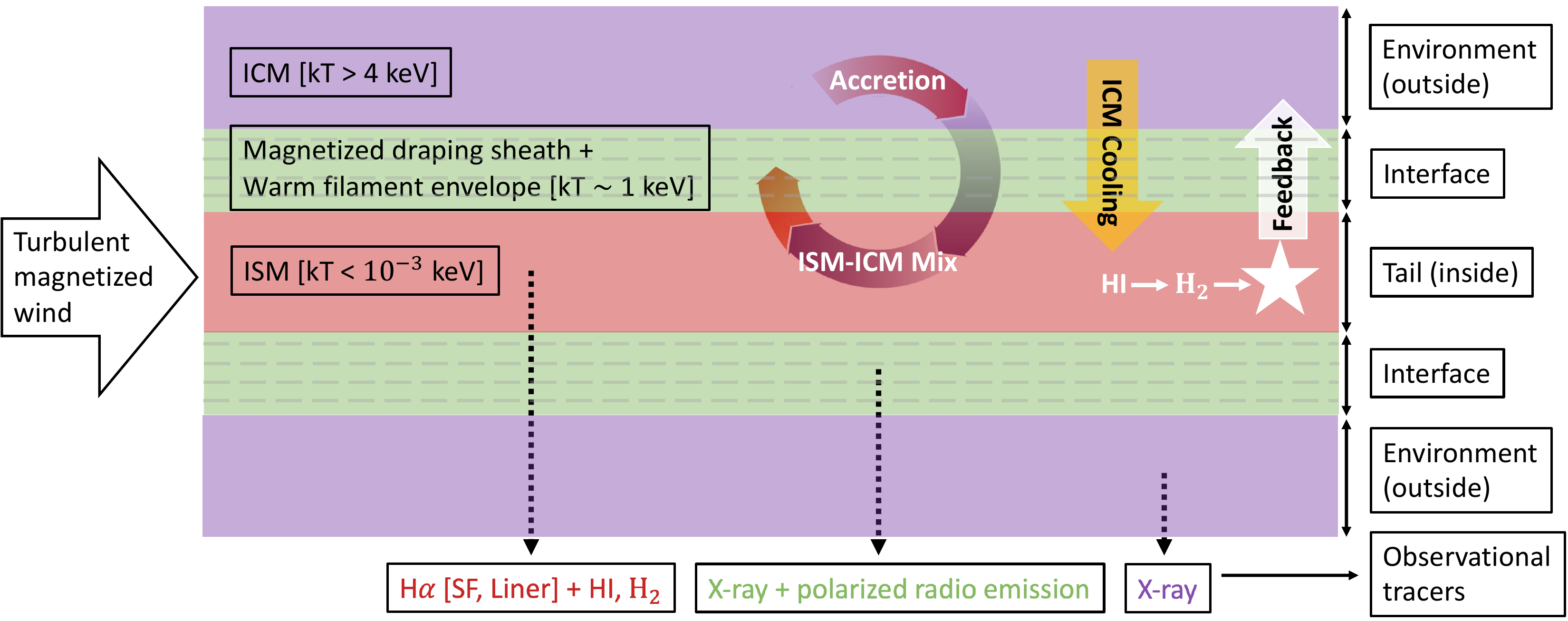}
\caption{Simplistic view of a section of the stripped tail depicting the holistic picture in ram-pressure stripped galaxies. The ICM is shown as outer layers (purple), the magnetic field and accreted/condensed ICM below (green with grey dotted lines indicating the highly aligned draped magnetic field), and the ISM consisting of different gas phases in the inner layer (red). While the ICM is accreted joining the ISM via mixing (circle arrow), ICM cooling is taking place outside-in (yellow arrow). Effective H\textsc{i} to H$_2$ conversion provides new stars that will provide feedback inside-out during stellar evolution and explosion processes (white arrow). \label{cartoon}}
\end{figure*}

In \citet{Mueller2021} we have performed state-of-the-art simulations with the moving mesh code AREPO \cite{Springel,Pakmor13,Pakmor15} of a galaxy-sized, cold-dense spherical cloud that interacts with a hot-diffuse ICM \citep[see also][]{Sparre} to model synchrotron characteristics. We adopt a scenario in which the electrons are accelerated by supernovae into the galaxies' disk and propagate into the surrounding interstellar medium. While the medium experiences ram-pressure stripping, the electrons are carried into the galaxies' tail and cool by synchrotron emission and inverse Compton interactions with the cosmic microwave background photons.
To grasp the essential physics the cloud size, wind field topologies, wind velocity, and cloud-wind density were adjusted to match the observations. The supersonic wind the cloud is facing has therefore been divided into three scenarios: 1) a purely hydrodynamical wind, 2) a wind composed of a homogeneous magnetic field, and 3) a turbulent magnetized wind. 
The density of the cloud is chosen to be a factor of $10^3$ higher than the wind and is in pressure equilibrium with the surroundings, and we adopt a weak magnetic field by assuming a thermal-to-magnetic pressure ratio of ten. To emulate the synchrotron characteristics we model the electron transport by identifying stripped gas cells as part of the galactic gas via Lagrangian trajectories. We follow the cooling of the electrons during the stripping process with a standard cooling function \citep[see Methods and Supplementary Information in][for any more details]{Mueller2021}. 

In the hydrodynamical case we found the Kelvin-Helmholtz instabilities to grow small perturbations at the wind-cloud interface and eventually disrupt the cloud. Therefore, in case of both magnetized winds, we found the formation of a magnetic draping sheath protecting the cloud from disruption forming magnetized filaments in the wake of the cloud (gaseous tail) \citep{Dursi,Berlok}. Here, case 3) was found to provide the best protection for the ram-pressure stripped, gaseous tail. While we found the gas cloud to sweep up magnetised ICM, building a draping layer around the head, a magnetised dense gaseous tail is formed via condensation of the ICM \citep{Sparre}. Thereby, the magnetic field is stretched and aligns with the tail (stripping direction). In addition, for large clouds (comparable to galaxy sizes), the cold phase in the ram-pressure stripped tail is found to survive and even increase due to gas accretion from the hot wind by mixing it into a layer with intermediate temperatures ($T\sim3\times10^5$\,K), from where it quickly cools and grows the cold tail \citep{Gronke,Li2020,Sparre}.

We further simulated synchrotron observables at 2.7\,GHz, comparable to the observation frequency of JO206, and found a gas cloud facing a turbulent magnetized wind to be the best fit \citep[see Fig. 5 in][]{Mueller2021}. Both, case 2) and 3) generally agree with the magnetic field findings for JO206, build an highly ordered magnetic field in the tail that can be explained by the magnetic draping of the magnetised ICM field and the accretion of gas from that sheath. However, the homogeneous wind-field naturally aligns with the tail and reaches, therefore, a non-physical high degree of polarization above 75\,\% while field reversals in the turbulent wind-field reduce the magnetic tension. In the latter case, the tail magnetic field becomes aligned and stretched with the accretion of mixed warm gas to the cold phase resembling the observation parameters of JO206 most accurately.

\section{Holistic picture}\label{sec:holipic}

The combined analysis of the different gas phases for a sub-sample of GASP galaxies and the unique polarization study of JO206, as well as the fact that radio tails are commonly found in ram-pressure stripped galaxies, puts forwards the idea that a highly-ordered magnetic field may be common and play a crucial role in regulating the star formation in the stripped tails. The following picture can be drawn (shown schematically in Fig. \ref{cartoon}).
We found the observed highly-ordered magnetic field far out into the tail of JO206 to be in excellent agreement with magneto-hydrodynamic simulations of a gas cloud exposed to turbulent ICM magnetic field. The galaxy-sized cloud was found to accrete hot, draped magnetised plasma that further condenses onto the outer layers of the tail (formation of a warm envelope shown with green layer in Fig. \ref{cartoon}). The adiabatic compression and shearing by the velocity difference of the cold tail and the wind thereafter can explain the observed high degree of fractional polarization and ordered magnetic field in JO206. We propose a formation of a magnetic draping layer at the galaxies' leading edges and shearing along the length of the tail which is indicated by the green layer in Fig. \ref{cartoon} and the dotted line as aligned magnetic field. In such a scenario, the galaxy facing the ram-pressure wind would be prevented from evaporating its cold ISM into the hot ICM (which is why the circle arrow stops inside the tail in Fig. \ref{cartoon}). Moreover, additional accretion of the cooled ICM from the outside-in (indicated by the yellow arrow in Fig. \ref{cartoon}) can provide new gas to form stars. The cooling timescales are still an open problem, however, given the physical context, non-equilibrium cooling processes are likely taking place (e.g., mixing). In order to investigate these non-linear processes, tailored numerical simulations are necessary.
Beside the JO206 observables, which already agree with such a scenario, we find more evidence in the multi-phase gas complementing this view.

The magnetic field configuration and star formation rate should depend on the amount of gas that is stripped from the disc and accreted from the hot ICM. The highly ordered magnetic field is expected to enter a phase where it protects the stripped gas from the disk into the tail from evaporation into the ICM by preventing heat and momentum exchange between the ISM and ICM. It is assumed that the accreted magnetised ICM cools while joining the cold-warm gaseous component of the ram-pressure stripped galaxy tails, possibly providing additional gas to form stars. 
Such a scenario cannot be probed by analysing only the magnetic field structure, but also by investigating the different gas phases. 
More explicitly, the hot ICM cooling along the tail during the accretion process should be observable in the X-ray band, explaining the extended X-ray emission found in different ram-pressure stripped tails (being part of the warm envelope shown in green in Fig. \ref{cartoon}). The thermal bremsstrahlung emission of the accreted gas could, in turn, be able to affect the ionized gas phases, as indicated by the excess in the ionization parameter ([OI]/H$\alpha$, indicated by the inner red layer in Fig. \ref{cartoon}).
Indeed a tentative connection between the thermal emission and the optical spectral properties in the stripped tails has been proposed in \citet{Campitiello2021} \footnote{A similar scenario was previously proposed also to explain the optical properties of the  cold filaments observed at the center of relaxed, cool core clusters \citep[e.g.,][]{Donald2012,Olivares2019}.}

Further cooling would provide additional H\textsc{i} joining the stripped gas from the disk. The magnetic pressure and tail structure possibly have an impact on the increased transformation from H\textsc{i} to H$_2$, therefore, the star-formation rates (shown by the white cartoon in Fig. \ref{cartoon}, where the white arrows indicate the effective transformation). 
In such a scenario, the magnetic field structure should be significantly ordered over larger time-scales ($\sim 10^8$\,yr) to prevent the tail gas from evaporation while the accreted ICM gas can effectively cool. Within the star-forming regions, freshly ejected electrons join the magnetic field component mainly caused by the ICM (and also provide feedback inside-out, both indicated by the white arrow in Fig. \ref{cartoon}) but won't disrupt it, otherwise the gas is expected to evaporate into the ICM, the in-situ star-formation will stop eventually and the galaxy evolves into a passive galaxy. 


\section{Current limitation and future perspective}

The recent results depict a close connection on a sub-kpc scale between the magnetic field and the ISM. Therefore, in order to investigate the physics of this link high-resolution, sub-arcsecond radio observations are necessary. 
This is a current limitation that could potentially be solved in the near future by the sub-arcsecond imaging made possible by the international LOFAR stations \citep[][]{Morabito_2021}. A second, crucial current limitation is the lack of candidates suitable for these studies, i.e., galaxies with both polarimetric and X-ray observations. In this regard, within GASP we are working to expand this sample. We have already been awarded time to observe deep polarization for four more galaxies -- JO204 and JW100 at the VLA, partly observed ($2-4$\,GHz C-array, ID: 21A-248, PI: Ancla M\"uller) -- JO147, partly observed ($1-3$\,GHz multiple array configurations, ID: C3357, PI: Ancla M\"uller) -- JO175 to be completed in Jan, 2022 ($1-3$\,GHz multiple array configurations, ID: C3357, PI: Ancla M\"uller) -- enabling us to probe the general existence of such a magnetic field configuration in star-forming, ram-pressure stripped tails.
In combination with the deep \textit{Chandra} observation (ID: 23610079, PI: Ignesti) of JO206 and its surrounding ICM we will be able to unfold the possible impact of the magnetic field onto the entire multi-phase gas.

Finally, the advent of deep radio surveys made possible by the new-generation facilities such as LOFAR, and SKA in the
future, provide us the opportunity to investigate the properties of jellyfish magnetic fields on larger samples of galaxies, and to evaluate the diversity of the cluster potential onto the connection of the magnetic field configuration and strength to the star-forming regions. Moreover, as outlined in \citep[][]{Ignesti_2020,Roberts2021,Roberts_2021b}, low-frequency LOFAR observations provide us an unique insight into the low-energy electrons travelling along the stripped tails, and the physical phenomena that drive their evolution. 
Future synergies between LOFAR, MeerkAT, VLA and, then SKA, will allow to extend this kind of analysis to larger samples of galaxies, thus deepening our knowledge of these fascinating systems.

\section{Conclusion}

We provide a summary of the recent results collected within the GASP collaboration including the analysis of the cold neutral and molecular gas, diffuse ionized gas tracers, hot gas, radio continuum and linear polarization. 
By combining the different observations we propose a holistic view of the evolution of galaxies experiencing ram-pressure stripping.
Here, we focus on the most massive galaxies and most elongated ram-pressure stripped tails forming stars in-situ analysed within GASP. Both the diverse observations and recent simulations of a gas cloud facing a turbulent, hot, magnetized ICM wind, agree with a scenario in which the galaxy accretes the hot plasma forming a magnetic draping layer at the leading edges and further along the tail preventing the stripped tail gas from evaporation while the ICM plasma becomes mixed with the ISM and cools from the outside-in. On larger timescales, the tail might be able to add mass from the ICM providing additional gas to possibly form new stars. We additionally argue that the formation of the magnetic draping sheath and the stars in the tail are a self-regulating process in which the outer layers will not survive due to a very efficient inside-out stellar feedback. We propose that the gas in the tails survives because of the combination of different effects: the galaxy mass, the galaxy infall velocity towards the cluster center, the pressure of the ICM surrounding the tails and the formation of the magnetic layer protecting this gas from evaporation. The further study of the phenomena presented in this work will add an important piece of information to our knowledge of the outcomes
of ram pressure stripping, as well as, more generally, to the scenario of draping, condensation and cooling of hot gas surrounding cold clouds that is fundamental in many astrophysical phenomena.


\authorcontributions{Conceptualization, An.M., B.P., and A.I.; investigation, An.M., B.P., and A.I.;  writing---original draft preparation, An.M and A.I.; writing---review and editing, An.M., A.I., B.P., A.M., M.R., and RJ.D. ; visualization, An.M., A.I., A.M., and M.R.; supervision, An.M., B.P., and RJ.D.; project administration, An.M. All authors have read and agreed to the published version of the manuscript.}

\funding{This project has received funding from the European Research  Council (ERC) under the European Union's Horizon 2020 research and innovation programme (grant agreement No. 833824). We acknowledge funding from the agreement ASI-INAF n.2017-14-H.0. (PI A. Moretti). We acknowledge financial support provided by SKA South-Africa and the South African National Research Foundation}

\dataavailability{The data presented in this study are available on request from the corresponding author.} 

\acknowledgments{Based on observations collected at the European Organization for Astronomical Research in the Southern Hemisphere under ESO programme 196.B-0578.
The authors are grateful to the whole GASP team for their contribution and useful discussions.}

\conflictsofinterest{The authors declare no conflict of interest.}


\end{paracol}
\reftitle{References}


\externalbibliography{yes}
\bibliography{literatur.bib}

\begin{thebibliography}{999}

\bibitem[{Noeske} \em{et~al.}(2007){Noeske}, {Weiner}, {Faber}, {Papovich},
  {Koo}, {Somerville}, {Bundy}, {Conselice}, {Newman}, {Schiminovich}, {Le
  Floc'h}, {Coil}, {Rieke}, {Lotz}, {Primack}, {Barmby}, {Cooper}, {Davis},
  {Ellis}, {Fazio}, {Guhathakurta}, {Huang}, {Kassin}, {Martin}, {Phillips},
  {Rich}, {Small}, {Willmer}, and {Wilson}]{Noeske2007}
{Noeske}, K.G.; {Weiner}, B.J.; {Faber}, S.M.; {Papovich}, C.; {Koo}, D.C.;
  {Somerville}, R.S.; {Bundy}, K.; {Conselice}, C.J.; {Newman}, J.A.;
  {Schiminovich}, D.; {Le Floc'h}, E.; {Coil}, A.L.; {Rieke}, G.H.; {Lotz},
  J.M.; {Primack}, J.R.; {Barmby}, P.; {Cooper}, M.C.; {Davis}, M.; {Ellis},
  R.S.; {Fazio}, G.G.; {Guhathakurta}, P.; {Huang}, J.; {Kassin}, S.A.;
  {Martin}, D.C.; {Phillips}, A.C.; {Rich}, R.M.; {Small}, T.A.; {Willmer},
  C.N.A.; {Wilson}, G.
\newblock {Star Formation in AEGIS Field Galaxies since z=1.1: The Dominance of
  Gradually Declining Star Formation, and the Main Sequence of Star-forming
  Galaxies}.
\newblock {\em Astrophysical Journal} {\bf 2007}, {\em 660},~L43--L46,
  \href{http://xxx.lanl.gov/abs/astro-ph/0701924}{{\normalfont
  [arXiv:astro-ph/astro-ph/0701924]}}.
\newblock
  doi:{\changeurlcolor{black}\href{https://doi.org/10.1086/517926}{\detokenize{10.1086/517926}}}.

\bibitem[Madau and Dickinson(2014)]{Madau2014}
Madau, P.; Dickinson, M.
\newblock Cosmic Star-Formation History.
\newblock {\em Annual Review of Astronomy and Astrophysics} {\bf 2014}, {\em
  52},~415–486.
\newblock
  doi:{\changeurlcolor{black}\href{https://doi.org/10.1146/annurev-astro-081811-125615}{\detokenize{10.1146/annurev-astro-081811-125615}}}.

\bibitem[{Boselli} and {Gavazzi}(2006)]{Boselli2006}
{Boselli}, A.; {Gavazzi}, G.
\newblock {Environmental Effects on Late-Type Galaxies in Nearby Clusters}.
\newblock {\em Publications of the Astronomical Society of the Pacific} {\bf
  2006}, {\em 118},~517--559,
  \href{http://xxx.lanl.gov/abs/astro-ph/0601108}{{\normalfont
  [arXiv:astro-ph/astro-ph/0601108]}}.
\newblock
  doi:{\changeurlcolor{black}\href{https://doi.org/10.1086/500691}{\detokenize{10.1086/500691}}}.

\bibitem[{Cayatte} \em{et~al.}(1990){Cayatte}, {van Gorkom}, {Balkowski}, and
  {Kotanyi}]{Cayatte}
{Cayatte}, V.; {van Gorkom}, J.H.; {Balkowski}, C.; {Kotanyi}, C.
\newblock {VLA Observations of Neutral Hydrogen in Virgo Cluster Galaxies. I.
  The Atlas}.
\newblock {\em Astronomical Journal} {\bf 1990}, {\em 100},~604.
\newblock
  doi:{\changeurlcolor{black}\href{https://doi.org/10.1086/115545}{\detokenize{10.1086/115545}}}.

\bibitem[Kenney \em{et~al.}(2004)Kenney, van Gorkom, and Vollmer]{Kenney2004}
Kenney, J.D.P.; van Gorkom, J.H.; Vollmer, B.
\newblock VLA HiObservations of Gas Stripping in the Virgo Cluster Spiral NGC
  4522.
\newblock {\em The Astronomical Journal} {\bf 2004}, {\em 127},~3361–3374.
\newblock
  doi:{\changeurlcolor{black}\href{https://doi.org/10.1086/420805}{\detokenize{10.1086/420805}}}.

\bibitem[{Chung} \em{et~al.}(2009){Chung}, {van Gorkom}, {Kenney}, {Crowl}, and
  {Vollmer}]{Chung09}
{Chung}, A.; {van Gorkom}, J.H.; {Kenney}, J.D.P.; {Crowl}, H.; {Vollmer}, B.
\newblock {VLA Imaging of Virgo Spirals in Atomic Gas (VIVA). I. The Atlas and
  the H I Properties}.
\newblock {\em Astrophysical Journal} {\bf 2009}, {\em 138},~1741--1816.
\newblock
  doi:{\changeurlcolor{black}\href{https://doi.org/10.1088/0004-6256/138/6/1741}{\detokenize{10.1088/0004-6256/138/6/1741}}}.

\bibitem[{Jaff{\'e}} \em{et~al.}(2015){Jaff{\'e}}, {Smith}, {Candlish},
  {Poggianti}, {Sheen}, and {Verheijen}]{Jaffe2015}
{Jaff{\'e}}, Y.L.; {Smith}, R.; {Candlish}, G.N.; {Poggianti}, B.M.; {Sheen},
  Y.K.; {Verheijen}, M.A.W.
\newblock {BUDHIES II: a phase-space view of H I gas stripping and star
  formation quenching in cluster galaxies}.
\newblock {\em Monthly Notices of the Royal Astronomical Society} {\bf 2015},
  {\em 448},~1715--1728,
  \href{http://xxx.lanl.gov/abs/1501.03819}{{\normalfont
  [arXiv:astro-ph.GA/1501.03819]}}.
\newblock
  doi:{\changeurlcolor{black}\href{https://doi.org/10.1093/mnras/stv100}{\detokenize{10.1093/mnras/stv100}}}.

\bibitem[{Scott} \em{et~al.}(2018){Scott}, {Brinks}, {Cortese}, {Boselli}, and
  {Bravo-Alfaro}]{Scott18}
{Scott}, T.C.; {Brinks}, E.; {Cortese}, L.; {Boselli}, A.; {Bravo-Alfaro}, H.
\newblock {Abell 1367: a high fraction of late-type galaxies displaying H I
  morphological and kinematic perturbations}.
\newblock {\em Monthly Notices of the Royal Astronomical Society} {\bf 2018},
  {\em 475},~4648--4669,
  \href{http://xxx.lanl.gov/abs/1801.03601}{{\normalfont
  [arXiv:astro-ph.GA/1801.03601]}}.
\newblock
  doi:{\changeurlcolor{black}\href{https://doi.org/10.1093/mnras/sty063}{\detokenize{10.1093/mnras/sty063}}}.

\bibitem[{Williams} and {Rood}(1987)]{William1987}
{Williams}, B.A.; {Rood}, H.J.
\newblock {Neutral Hydrogen in Compact Groups of Galaxies}.
\newblock {\em Astrophysical Journal} {\bf 1987}, {\em 63},~265.
\newblock
  doi:{\changeurlcolor{black}\href{https://doi.org/10.1086/191165}{\detokenize{10.1086/191165}}}.

\bibitem[{Verdes-Montenegro} \em{et~al.}(2001){Verdes-Montenegro}, {Yun},
  {Williams}, {Huchtmeier}, {Del Olmo}, and {Perea}]{Montenegro2001}
{Verdes-Montenegro}, L.; {Yun}, M.S.; {Williams}, B.A.; {Huchtmeier}, W.K.;
  {Del Olmo}, A.; {Perea}, J.
\newblock {Where is the neutral atomic gas in Hickson groups?}
\newblock {\em Astronomy \& Astrophysics} {\bf 2001}, {\em 377},~812--826,
  \href{http://xxx.lanl.gov/abs/astro-ph/0108223}{{\normalfont
  [arXiv:astro-ph/astro-ph/0108223]}}.
\newblock
  doi:{\changeurlcolor{black}\href{https://doi.org/10.1051/0004-6361:20011127}{\detokenize{10.1051/0004-6361:20011127}}}.

\bibitem[{Serra} \em{et~al.}(2013){Serra}, {Koribalski}, {Duc}, {Oosterloo},
  {McDermid}, {Michel-Dansac}, {Emsellem}, {Cuillandre}, {Alatalo}, {Blitz},
  {Bois}, {Bournaud}, {Bureau}, {Cappellari}, {Crocker}, {Davies}, {Davis}, {de
  Zeeuw}, {Khochfar}, {Krajnovi{\'c}}, {Kuntschner}, {Lablanche}, {Morganti},
  {Naab}, {Sarzi}, {Scott}, {Weijmans}, and {Young}]{Serra2013}
{Serra}, P.; {Koribalski}, B.; {Duc}, P.A.; {Oosterloo}, T.; {McDermid}, R.M.;
  {Michel-Dansac}, L.; {Emsellem}, E.; {Cuillandre}, J.C.; {Alatalo}, K.;
  {Blitz}, L.; {Bois}, M.; {Bournaud}, F.; {Bureau}, M.; {Cappellari}, M.;
  {Crocker}, A.F.; {Davies}, R.L.; {Davis}, T.A.; {de Zeeuw}, P.T.; {Khochfar},
  S.; {Krajnovi{\'c}}, D.; {Kuntschner}, H.; {Lablanche}, P.Y.; {Morganti}, R.;
  {Naab}, T.; {Sarzi}, M.; {Scott}, N.; {Weijmans}, A.M.; {Young}, L.M.
\newblock {Discovery of a giant HI tail in the galaxy group HCG 44}.
\newblock {\em Monthly Notices of the Royal Astronomical Society} {\bf 2013},
  {\em 428},~370--380,  \href{http://xxx.lanl.gov/abs/1209.4107}{{\normalfont
  [arXiv:astro-ph.CO/1209.4107]}}.
\newblock
  doi:{\changeurlcolor{black}\href{https://doi.org/10.1093/mnras/sts033}{\detokenize{10.1093/mnras/sts033}}}.

\bibitem[{Bellhouse} \em{et~al.}(2017){Bellhouse}, {Jaff{\'e}}, {Hau}, {McGee},
  {Poggianti}, {Moretti}, {Gullieuszik}, {Bettoni}, {Fasano}, {D'Onofrio},
  {Fritz}, {Omizzolo}, {Sheen}, and {Vulcani}]{Bellhouse2017}
{Bellhouse}, C.; {Jaff{\'e}}, Y.L.; {Hau}, G.K.T.; {McGee}, S.L.; {Poggianti},
  B.M.; {Moretti}, A.; {Gullieuszik}, M.; {Bettoni}, D.; {Fasano}, G.;
  {D'Onofrio}, M.; {Fritz}, J.; {Omizzolo}, A.; {Sheen}, Y.K.; {Vulcani}, B.
\newblock {GASP. II. A MUSE View of Extreme Ram-Pressure Stripping along the
  Line of Sight: Kinematics of the Jellyfish Galaxy JO201}.
\newblock {\em Astrophysical Journal} {\bf 2017}, {\em 844},~49,
  \href{http://xxx.lanl.gov/abs/1704.05087}{{\normalfont
  [arXiv:astro-ph.GA/1704.05087]}}.
\newblock
  doi:{\changeurlcolor{black}\href{https://doi.org/10.3847/1538-4357/aa7875}{\detokenize{10.3847/1538-4357/aa7875}}}.

\bibitem[{Boselli} \em{et~al.}(2021){Boselli}, {Fossati}, and {Sun}]{Boselli21}
{Boselli}, A.; {Fossati}, M.; {Sun}, M.
\newblock {Ram Pressure Stripping in High-Density Environments}.
\newblock {\em arXiv e-prints} {\bf 2021}, p. arXiv:2109.13614,
  \href{http://xxx.lanl.gov/abs/2109.13614}{{\normalfont
  [arXiv:astro-ph.GA/2109.13614]}}.

\bibitem[{Cortese} \em{et~al.}(2007){Cortese}, {Marcillac}, {Richard},
  {Bravo-Alfaro}, {Kneib}, {Rieke}, {Covone}, {Egami}, {Rigby}, {Czoske}, and
  {Davies}]{Cortese07}
{Cortese}, L.; {Marcillac}, D.; {Richard}, J.; {Bravo-Alfaro}, H.; {Kneib},
  J.P.; {Rieke}, G.; {Covone}, G.; {Egami}, E.; {Rigby}, J.; {Czoske}, O.;
  {Davies}, J.
\newblock {The strong transformation of spiral galaxies infalling into massive
  clusters at z \raisebox{-0.5ex}\textasciitilde 0.2}.
\newblock {\em Monthly Notices of the Royal Astronomical Society} {\bf 2007},
  {\em 376},~157--172,
  \href{http://xxx.lanl.gov/abs/astro-ph/0703012}{{\normalfont
  [arXiv:astro-ph/astro-ph/0703012]}}.
\newblock
  doi:{\changeurlcolor{black}\href{https://doi.org/10.1111/j.1365-2966.2006.11369.x}{\detokenize{10.1111/j.1365-2966.2006.11369.x}}}.

\bibitem[{Poggianti} \em{et~al.}(2019){Poggianti}, {Gullieuszik}, {Tonnesen},
  {Moretti}, {Vulcani}, {Radovich}, {Jaff{\'e}}, {Fritz}, {Bettoni},
  {Franchetto}, {Fasano}, {Bellhouse}, and {Omizzolo}]{Poggianti2019b}
{Poggianti}, B.M.; {Gullieuszik}, M.; {Tonnesen}, S.; {Moretti}, A.; {Vulcani},
  B.; {Radovich}, M.; {Jaff{\'e}}, Y.; {Fritz}, J.; {Bettoni}, D.;
  {Franchetto}, A.; {Fasano}, G.; {Bellhouse}, C.; {Omizzolo}, A.
\newblock {GASP XIII. Star formation in gas outside galaxies}.
\newblock {\em Monthly Notices of the Royal Astronomical Society} {\bf 2019},
  {\em 482},~4466--4502,
  \href{http://xxx.lanl.gov/abs/1811.00823}{{\normalfont
  [arXiv:astro-ph.GA/1811.00823]}}.
\newblock
  doi:{\changeurlcolor{black}\href{https://doi.org/10.1093/mnras/sty2999}{\detokenize{10.1093/mnras/sty2999}}}.

\bibitem[{Crutcher}(2012)]{Crutcher2012}
{Crutcher}, R.M.
\newblock {Magnetic Fields in Molecular Clouds}.
\newblock {\em Annual Review of Astronomy and Astrophysics} {\bf 2012}, {\em
  50},~29--63.
\newblock
  doi:{\changeurlcolor{black}\href{https://doi.org/10.1146/annurev-astro-081811-125514}{\detokenize{10.1146/annurev-astro-081811-125514}}}.

\bibitem[{Heesen} \em{et~al.}(2011){Heesen}, {Beck}, {Krause}, and
  {Dettmar}]{Heesen11}
{Heesen}, V.; {Beck}, R.; {Krause}, M.; {Dettmar}, R.J.
\newblock {Cosmic rays and the magnetic field in the nearby starburst galaxy
  NGC 253 III. Helical magnetic fields in the nuclear outflow}.
\newblock {\em Astronomy and Astrophysics} {\bf 2011}, {\em 535},~A79,
  \href{http://xxx.lanl.gov/abs/1109.0255}{{\normalfont
  [arXiv:astro-ph.CO/1109.0255]}}.
\newblock
  doi:{\changeurlcolor{black}\href{https://doi.org/10.1051/0004-6361/201117618}{\detokenize{10.1051/0004-6361/201117618}}}.

\bibitem[{Beck} and {Wielebinski}(2013)]{Beck2013}
{Beck}, R.; {Wielebinski}, R., {Magnetic Fields in Galaxies}.
\newblock In {\em Planets, Stars and Stellar Systems. Volume 5: Galactic
  Structure and Stellar Populations}; Springer,  2013; Vol.~5, p. 641.
\newblock
  doi:{\changeurlcolor{black}\href{https://doi.org/10.1007/978-94-007-5612-0\_13}{\detokenize{10.1007/978-94-007-5612-0\_13}}}.

\bibitem[{Li} and {Henning}(2011)]{Li2011}
{Li}, H.B.; {Henning}, T.
\newblock {The alignment of molecular cloud magnetic fields with the spiral
  arms in M33}.
\newblock {\em Nature} {\bf 2011}, {\em 479},~499--501,
  \href{http://xxx.lanl.gov/abs/1111.2745}{{\normalfont
  [arXiv:astro-ph.GA/1111.2745]}}.
\newblock
  doi:{\changeurlcolor{black}\href{https://doi.org/10.1038/nature10551}{\detokenize{10.1038/nature10551}}}.

\bibitem[{Gavazzi} \em{et~al.}(1995){Gavazzi}, {Contursi}, {Carrasco},
  {Boselli}, {Kennicutt}, {Scodeggio}, and {Jaffe}]{Gavazzi1995}
{Gavazzi}, G.; {Contursi}, A.; {Carrasco}, L.; {Boselli}, A.; {Kennicutt}, R.;
  {Scodeggio}, M.; {Jaffe}, W.
\newblock {The radio and optical structure of three peculiar galaxies in A
  1367.}
\newblock {\em Astronomy and Astrophysics} {\bf 1995}, {\em 304},~325.

\bibitem[{Vollmer} \em{et~al.}(2010){Vollmer}, {Soida}, {Chung}, {Beck},
  {Urbanik}, {Chy{\.z}y}, {Otmianowska-Mazur}, and {van Gorkom}]{Vollmer2010}
{Vollmer}, B.; {Soida}, M.; {Chung}, A.; {Beck}, R.; {Urbanik}, M.;
  {Chy{\.z}y}, K.T.; {Otmianowska-Mazur}, K.; {van Gorkom}, J.H.
\newblock {The influence of the cluster environment on the large-scale radio
  continuum emission of 8 Virgo cluster spirals}.
\newblock {\em Astronomy \& Astrophysics} {\bf 2010}, {\em 512},~A36,
  \href{http://xxx.lanl.gov/abs/1001.3597}{{\normalfont
  [arXiv:astro-ph.CO/1001.3597]}}.
\newblock
  doi:{\changeurlcolor{black}\href{https://doi.org/10.1051/0004-6361/200913591}{\detokenize{10.1051/0004-6361/200913591}}}.

\bibitem[{Chen} \em{et~al.}(2020){Chen}, {Sun}, {Yagi}, {Bravo-Alfaro},
  {Brinks}, {Kenney}, {Combes}, {Sivanandam}, {Jachym}, {Fossati}, {Gavazzi},
  {Boselli}, {Nulsen}, {Sarazin}, {Ge}, {Yoshida}, and {Roediger}]{Chen}
{Chen}, H.; {Sun}, M.; {Yagi}, M.; {Bravo-Alfaro}, H.; {Brinks}, E.; {Kenney},
  J.; {Combes}, F.; {Sivanandam}, S.; {Jachym}, P.; {Fossati}, M.; {Gavazzi},
  G.; {Boselli}, A.; {Nulsen}, P.; {Sarazin}, C.; {Ge}, C.; {Yoshida}, M.;
  {Roediger}, E.
\newblock {The ram pressure stripped radio tails of galaxies in the Coma
  cluster}.
\newblock {\em Monthly Notices of the Royal Astronomical Society} {\bf 2020},
  {\em 496},~4654--4673,
  \href{http://xxx.lanl.gov/abs/2004.06743}{{\normalfont
  [arXiv:astro-ph.GA/2004.06743]}}.
\newblock
  doi:{\changeurlcolor{black}\href{https://doi.org/10.1093/mnras/staa1868}{\detokenize{10.1093/mnras/staa1868}}}.

\bibitem[{M{\"u}ller} \em{et~al.}(2021){M{\"u}ller}, {Poggianti}, {Pfrommer},
  {Adebahr}, {Serra}, {Ignesti}, {Sparre}, {Gitti}, {Dettmar}, {Vulcani}, and
  {Moretti}]{Mueller2021}
{M{\"u}ller}, A.; {Poggianti}, B.M.; {Pfrommer}, C.; {Adebahr}, B.; {Serra},
  P.; {Ignesti}, A.; {Sparre}, M.; {Gitti}, M.; {Dettmar}, R.J.; {Vulcani}, B.;
  {Moretti}, A.
\newblock {Highly ordered magnetic fields in the tail of the jellyfish galaxy
  JO206}.
\newblock {\em Nature Astronomy} {\bf 2021}, {\em 5},~159--168,
  \href{http://xxx.lanl.gov/abs/2009.13287}{{\normalfont
  [arXiv:astro-ph.GA/2009.13287]}}.
\newblock
  doi:{\changeurlcolor{black}\href{https://doi.org/10.1038/s41550-020-01234-7}{\detokenize{10.1038/s41550-020-01234-7}}}.

\bibitem[{Ignesti} \em{et~al.}(2021){Ignesti}, {Vulcani}, {Poggianti},
  {Paladino}, {Shimwell}, {Healy}, {Gitti}, {Bacchini}, {Moretti}, {Radovich},
  {van Weeren}, {Roberts}, {Botteon}, {M{\"u}ller}, {McGee}, {Fritz},
  {Tom{\v{c}}i{\'c}}, {Werle}, {Mingozzi}, {Gullieuszik}, and
  {Verheijen}]{Ignesti21}
{Ignesti}, A.; {Vulcani}, B.; {Poggianti}, B.M.; {Paladino}, R.; {Shimwell},
  T.; {Healy}, J.; {Gitti}, M.; {Bacchini}, C.; {Moretti}, A.; {Radovich}, M.;
  {van Weeren}, R.J.; {Roberts}, I.D.; {Botteon}, A.; {M{\"u}ller}, A.;
  {McGee}, S.; {Fritz}, J.; {Tom{\v{c}}i{\'c}}, N.; {Werle}, A.; {Mingozzi},
  M.; {Gullieuszik}, M.; {Verheijen}, M.
\newblock {GASP XXXVIII: The LOFAR-MeerKAT-VLA view on the non-thermal side of
  a jellyfish galaxy}.
\newblock {\em arXiv e-prints} {\bf 2021}, p. arXiv:2110.12719,
  \href{http://xxx.lanl.gov/abs/2110.12719}{{\normalfont
  [arXiv:astro-ph.GA/2110.12719]}}.

\bibitem[Poggianti \em{et~al.}(2019)Poggianti, Ignesti, Gitti, Wolter,
  Brighenti, Biviano, George, Vulcani, Gullieuszik, Moretti, and
  et~al.]{Poggianti2019a}
Poggianti, B.M.; Ignesti, A.; Gitti, M.; Wolter, A.; Brighenti, F.; Biviano,
  A.; George, K.; Vulcani, B.; Gullieuszik, M.; Moretti, A.; et~al..
\newblock GASP XXIII: A Jellyfish Galaxy as an Astrophysical Laboratory of the
  Baryonic Cycle.
\newblock {\em The Astrophysical Journal} {\bf 2019}, {\em 887},~155.
\newblock
  doi:{\changeurlcolor{black}\href{https://doi.org/10.3847/1538-4357/ab5224}{\detokenize{10.3847/1538-4357/ab5224}}}.

\bibitem[{Shimwell} \em{et~al.}(2017){Shimwell}, {R{\"o}ttgering}, {Best},
  {Williams}, {Dijkema}, {de Gasperin}, {Hardcastle}, {Heald}, {Hoang},
  {Horneffer}, {Intema}, {Mahony}, {Mandal}, {Mechev}, {Morabito}, {Oonk},
  {Rafferty}, {Retana-Montenegro}, {Sabater}, {Tasse}, {van Weeren},
  {Br{\"u}ggen}, {Brunetti}, {Chy{\.z}y}, {Conway}, {Haverkorn}, {Jackson},
  {Jarvis}, {McKean}, {Miley}, {Morganti}, {White}, {Wise}, {van Bemmel},
  {Beck}, {Brienza}, {Bonafede}, {Calistro Rivera}, {Cassano}, {Clarke},
  {Cseh}, {Deller}, {Drabent}, {van Driel}, {Engels}, {Falcke}, {Ferrari},
  {Fr{\"o}hlich}, {Garrett}, {Harwood}, {Heesen}, {Hoeft}, {Horellou},
  {Israel}, {Kapi{\'n}ska}, {Kunert-Bajraszewska}, {McKay}, {Mohan},
  {Orr{\'u}}, {Pizzo}, {Prandoni}, {Schwarz}, {Shulevski}, {Sipior}, {Smith},
  {Sridhar}, {Steinmetz}, {Stroe}, {Varenius}, {van der Werf}, {Zensus}, and
  {Zwart}]{Shimwell_2017}
{Shimwell}, T.W.; {R{\"o}ttgering}, H.J.A.; {Best}, P.N.; {Williams}, W.L.;
  {Dijkema}, T.J.; {de Gasperin}, F.; {Hardcastle}, M.J.; {Heald}, G.H.;
  {Hoang}, D.N.; {Horneffer}, A.; {Intema}, H.; {Mahony}, E.K.; {Mandal}, S.;
  {Mechev}, A.P.; {Morabito}, L.; {Oonk}, J.B.R.; {Rafferty}, D.;
  {Retana-Montenegro}, E.; {Sabater}, J.; {Tasse}, C.; {van Weeren}, R.J.;
  {Br{\"u}ggen}, M.; {Brunetti}, G.; {Chy{\.z}y}, K.T.; {Conway}, J.E.;
  {Haverkorn}, M.; {Jackson}, N.; {Jarvis}, M.J.; {McKean}, J.P.; {Miley},
  G.K.; {Morganti}, R.; {White}, G.J.; {Wise}, M.W.; {van Bemmel}, I.M.;
  {Beck}, R.; {Brienza}, M.; {Bonafede}, A.; {Calistro Rivera}, G.; {Cassano},
  R.; {Clarke}, A.O.; {Cseh}, D.; {Deller}, A.; {Drabent}, A.; {van Driel}, W.;
  {Engels}, D.; {Falcke}, H.; {Ferrari}, C.; {Fr{\"o}hlich}, S.; {Garrett},
  M.A.; {Harwood}, J.J.; {Heesen}, V.; {Hoeft}, M.; {Horellou}, C.; {Israel},
  F.P.; {Kapi{\'n}ska}, A.D.; {Kunert-Bajraszewska}, M.; {McKay}, D.J.;
  {Mohan}, N.R.; {Orr{\'u}}, E.; {Pizzo}, R.F.; {Prandoni}, I.; {Schwarz},
  D.J.; {Shulevski}, A.; {Sipior}, M.; {Smith}, D.J.B.; {Sridhar}, S.S.;
  {Steinmetz}, M.; {Stroe}, A.; {Varenius}, E.; {van der Werf}, P.P.; {Zensus},
  J.A.; {Zwart}, J.T.L.
\newblock {The LOFAR Two-metre Sky Survey. I. Survey description and
  preliminary data release}.
\newblock {\em Astronomy and Astrophysics} {\bf 2017}, {\em 598},~A104,
  \href{http://xxx.lanl.gov/abs/1611.02700}{{\normalfont
  [arXiv:astro-ph.IM/1611.02700]}}.
\newblock
  doi:{\changeurlcolor{black}\href{https://doi.org/10.1051/0004-6361/201629313}{\detokenize{10.1051/0004-6361/201629313}}}.

\bibitem[Roberts \em{et~al.}(2021)Roberts, van Weeren, McGee, Botteon, Drabent,
  Ignesti, Rottgering, Shimwell, and Tasse]{Roberts2021}
Roberts, I.D.; van Weeren, R.J.; McGee, S.L.; Botteon, A.; Drabent, A.;
  Ignesti, A.; Rottgering, H.J.A.; Shimwell, T.W.; Tasse, C.
\newblock LoTSS jellyfish galaxies.
\newblock {\em Astronomy \& Astrophysics} {\bf 2021}, {\em 650},~A111.
\newblock
  doi:{\changeurlcolor{black}\href{https://doi.org/10.1051/0004-6361/202140784}{\detokenize{10.1051/0004-6361/202140784}}}.

\bibitem[{Roberts} \em{et~al.}(2021){Roberts}, {van Weeren}, {McGee},
  {Botteon}, {Ignesti}, and {Rottgering}]{Roberts_2021b}
{Roberts}, I.D.; {van Weeren}, R.J.; {McGee}, S.L.; {Botteon}, A.; {Ignesti},
  A.; {Rottgering}, H.J.A.
\newblock {LoTSS jellyfish galaxies: II. Ram pressure stripping in groups
  versus clusters}.
\newblock {\em arXiv e-prints} {\bf 2021}, p. arXiv:2106.06315,
  \href{http://xxx.lanl.gov/abs/2106.06315}{{\normalfont
  [arXiv:astro-ph.GA/2106.06315]}}.

\bibitem[{Poggianti} \em{et~al.}(2017){Poggianti}, {Moretti}, {Gullieuszik},
  {Fritz}, {Jaff{\'e}}, {Bettoni}, {Fasano}, {Bellhouse}, {Hau}, {Vulcani},
  {Biviano}, {Omizzolo}, {Paccagnella}, {D'Onofrio}, {Cava}, {Sheen}, {Couch},
  and {Owers}]{Poggianti2017}
{Poggianti}, B.M.; {Moretti}, A.; {Gullieuszik}, M.; {Fritz}, J.; {Jaff{\'e}},
  Y.; {Bettoni}, D.; {Fasano}, G.; {Bellhouse}, C.; {Hau}, G.; {Vulcani}, B.;
  {Biviano}, A.; {Omizzolo}, A.; {Paccagnella}, A.; {D'Onofrio}, M.; {Cava},
  A.; {Sheen}, Y.K.; {Couch}, W.; {Owers}, M.
\newblock {GASP. I. Gas Stripping Phenomena in Galaxies with MUSE}.
\newblock {\em Astrophysical Journal} {\bf 2017}, {\em 844},~48,
  \href{http://xxx.lanl.gov/abs/1704.05086}{{\normalfont
  [arXiv:astro-ph.GA/1704.05086]}}.
\newblock
  doi:{\changeurlcolor{black}\href{https://doi.org/10.3847/1538-4357/aa78ed}{\detokenize{10.3847/1538-4357/aa78ed}}}.

\bibitem[{Poggianti} \em{et~al.}(2016){Poggianti}, {Fasano}, {Omizzolo},
  {Gullieuszik}, {Bettoni}, {Moretti}, {Paccagnella}, {Jaff{\'e}}, {Vulcani},
  {Fritz}, {Couch}, and {D'Onofrio}]{Poggianti2016}
{Poggianti}, B.M.; {Fasano}, G.; {Omizzolo}, A.; {Gullieuszik}, M.; {Bettoni},
  D.; {Moretti}, A.; {Paccagnella}, A.; {Jaff{\'e}}, Y.L.; {Vulcani}, B.;
  {Fritz}, J.; {Couch}, W.; {D'Onofrio}, M.
\newblock {Jellyfish Galaxy Candidates at Low Redshift}.
\newblock {\em Astronomical Journal} {\bf 2016}, {\em 151},~78,
  \href{http://xxx.lanl.gov/abs/1504.07105}{{\normalfont
  [arXiv:astro-ph.GA/1504.07105]}}.
\newblock
  doi:{\changeurlcolor{black}\href{https://doi.org/10.3847/0004-6256/151/3/78}{\detokenize{10.3847/0004-6256/151/3/78}}}.

\bibitem[{Ramatsoku} \em{et~al.}(2019){Ramatsoku}, {Serra}, {Poggianti},
  {Moretti}, {Gullieuszik}, {Bettoni}, {Deb}, {Fritz}, {van Gorkom},
  {Jaff{\'e}}, {Tonnesen}, {Verheijen}, {Vulcani}, {Hugo}, {J{\'o}zsa},
  {Maccagni}, {Makhathini}, {Ramaila}, {Smirnov}, and {Thorat}]{Ramatsoku2019}
{Ramatsoku}, M.; {Serra}, P.; {Poggianti}, B.M.; {Moretti}, A.; {Gullieuszik},
  M.; {Bettoni}, D.; {Deb}, T.; {Fritz}, J.; {van Gorkom}, J.H.; {Jaff{\'e}},
  Y.L.; {Tonnesen}, S.; {Verheijen}, M.A.W.; {Vulcani}, B.; {Hugo}, B.;
  {J{\'o}zsa}, G.I.G.; {Maccagni}, F.M.; {Makhathini}, S.; {Ramaila}, A.;
  {Smirnov}, O.; {Thorat}, K.
\newblock {GASP - XVII. H I imaging of the jellyfish galaxy JO206: gas
  stripping and enhanced star formation}.
\newblock {\em Monthly Notices of the Royal Astronomical Society} {\bf 2019},
  {\em 487},~4580--4591,
  \href{http://xxx.lanl.gov/abs/1906.03686}{{\normalfont
  [arXiv:astro-ph.GA/1906.03686]}}.
\newblock
  doi:{\changeurlcolor{black}\href{https://doi.org/10.1093/mnras/stz1609}{\detokenize{10.1093/mnras/stz1609}}}.

\bibitem[{Ramatsoku} \em{et~al.}(2020){Ramatsoku}, {Serra}, {Poggianti},
  {Moretti}, {Gullieuszik}, {Bettoni}, {Deb}, {Franchetto}, {van Gorkom},
  {Jaff{\'e}}, {Tonnesen}, {Verheijen}, {Vulcani}, {Andati}, {de Blok},
  {J{\'o}zsa}, {Kamphuis}, {Kleiner}, {Maccagni}, {Makhathini}, {Moln{\'a}r},
  {Ramaila}, {Smirnov}, and {Thorat}]{Ramatsoku2020}
{Ramatsoku}, M.; {Serra}, P.; {Poggianti}, B.M.; {Moretti}, A.; {Gullieuszik},
  M.; {Bettoni}, D.; {Deb}, T.; {Franchetto}, A.; {van Gorkom}, J.H.;
  {Jaff{\'e}}, Y.; {Tonnesen}, S.; {Verheijen}, M.A.W.; {Vulcani}, B.;
  {Andati}, L.A.L.; {de Blok}, E.; {J{\'o}zsa}, G.I.G.; {Kamphuis}, P.;
  {Kleiner}, D.; {Maccagni}, F.M.; {Makhathini}, S.; {Moln{\'a}r}, D.C.;
  {Ramaila}, A.J.T.; {Smirnov}, O.; {Thorat}, K.
\newblock {GASP. XXVI. HI gas in jellyfish galaxies: The case of JO201 and
  JO206}.
\newblock {\em Astronomy \& Astrophysics} {\bf 2020}, {\em 640},~A22,
  \href{http://xxx.lanl.gov/abs/2006.11543}{{\normalfont
  [arXiv:astro-ph.GA/2006.11543]}}.
\newblock
  doi:{\changeurlcolor{black}\href{https://doi.org/10.1051/0004-6361/202037759}{\detokenize{10.1051/0004-6361/202037759}}}.

\bibitem[{Deb} \em{et~al.}(2020){Deb}, {Verheijen}, {Gullieuszik}, {Poggianti},
  {van Gorkom}, {Ramatsoku}, {Serra}, {Moretti}, {Vulcani}, {Bettoni},
  {Jaff{\'e}}, {Tonnesen}, and {Fritz}]{Deb2020}
{Deb}, T.; {Verheijen}, M.A.W.; {Gullieuszik}, M.; {Poggianti}, B.M.; {van
  Gorkom}, J.H.; {Ramatsoku}, M.; {Serra}, P.; {Moretti}, A.; {Vulcani}, B.;
  {Bettoni}, D.; {Jaff{\'e}}, L.Y.; {Tonnesen}, S.; {Fritz}, J.
\newblock {GASP XXV: neutral hydrogen gas in the striking jellyfish galaxy
  JO204}.
\newblock {\em Monthly Notices of the Royal Astronomical Society} {\bf 2020},
  {\em 494},~5029--5043,
  \href{http://xxx.lanl.gov/abs/2004.04754}{{\normalfont
  [arXiv:astro-ph.GA/2004.04754]}}.
\newblock
  doi:{\changeurlcolor{black}\href{https://doi.org/10.1093/mnras/staa968}{\detokenize{10.1093/mnras/staa968}}}.

\bibitem[{Moretti} \em{et~al.}(2018){Moretti}, {Paladino}, {Poggianti},
  {D'Onofrio}, {Bettoni}, {Gullieuszik}, {Jaff{\'e}}, {Vulcani}, {Fasano},
  {Fritz}, and {Torstensson}]{Moretti2018}
{Moretti}, A.; {Paladino}, R.; {Poggianti}, B.M.; {D'Onofrio}, M.; {Bettoni},
  D.; {Gullieuszik}, M.; {Jaff{\'e}}, Y.L.; {Vulcani}, B.; {Fasano}, G.;
  {Fritz}, J.; {Torstensson}, K.
\newblock {GASP - X. APEX observations of molecular gas in the discs and in the
  tails of ram-pressure stripped galaxies}.
\newblock {\em Monthly Notices of the Royal Astronomical Society} {\bf 2018},
  {\em 480},~2508--2520,
  \href{http://xxx.lanl.gov/abs/1803.06183}{{\normalfont
  [arXiv:astro-ph.GA/1803.06183]}}.
\newblock
  doi:{\changeurlcolor{black}\href{https://doi.org/10.1093/mnras/sty2021}{\detokenize{10.1093/mnras/sty2021}}}.

\bibitem[{Moretti} \em{et~al.}(2020){Moretti}, {Paladino}, {Poggianti},
  {Serra}, {Ramatsoku}, {Franchetto}, {Deb}, {Gullieuszik},
  {Tomi{\v{c}}i{\'c}}, {Mingozzi}, {Vulcani}, {Radovich}, {Bettoni}, and
  {Fritz}]{Moretti2020}
{Moretti}, A.; {Paladino}, R.; {Poggianti}, B.M.; {Serra}, P.; {Ramatsoku}, M.;
  {Franchetto}, A.; {Deb}, T.; {Gullieuszik}, M.; {Tomi{\v{c}}i{\'c}}, N.;
  {Mingozzi}, M.; {Vulcani}, B.; {Radovich}, M.; {Bettoni}, D.; {Fritz}, J.
\newblock {The High Molecular Gas Content, and the Efficient Conversion of
  Neutral into Molecular Gas, in Jellyfish Galaxies}.
\newblock {\em Astrophysical Journal} {\bf 2020}, {\em 897},~L30,
  \href{http://xxx.lanl.gov/abs/2006.13612}{{\normalfont
  [arXiv:astro-ph.GA/2006.13612]}}.
\newblock
  doi:{\changeurlcolor{black}\href{https://doi.org/10.3847/2041-8213/ab9f3b}{\detokenize{10.3847/2041-8213/ab9f3b}}}.

\bibitem[{Campitiello} \em{et~al.}(2021){Campitiello}, {Ignesti}, {Gitti},
  {Brighenti}, {Radovich}, {Wolter}, {Tomi{\v{c}}i{\'c}}, {Bellhouse},
  {Poggianti}, {Moretti}, {Vulcani}, {Jaff{\'e}}, {Paladino}, {M{\"u}ller},
  {Fritz}, {Louren{\c{c}}o}, and {Gullieuszik}]{Campitiello2021}
{Campitiello}, M.G.; {Ignesti}, A.; {Gitti}, M.; {Brighenti}, F.; {Radovich},
  M.; {Wolter}, A.; {Tomi{\v{c}}i{\'c}}, N.; {Bellhouse}, C.; {Poggianti},
  B.M.; {Moretti}, A.; {Vulcani}, B.; {Jaff{\'e}}, Y.L.; {Paladino}, R.;
  {M{\"u}ller}, A.; {Fritz}, J.; {Louren{\c{c}}o}, A.C.C.; {Gullieuszik}, M.
\newblock {GASP XXXIV: Unfolding the Thermal Side of Ram Pressure Stripping in
  the Jellyfish Galaxy JO201}.
\newblock {\em Astrophysical Journal} {\bf 2021}, {\em 911},~144,
  \href{http://xxx.lanl.gov/abs/2103.03147}{{\normalfont
  [arXiv:astro-ph.GA/2103.03147]}}.
\newblock
  doi:{\changeurlcolor{black}\href{https://doi.org/10.3847/1538-4357/abec82}{\detokenize{10.3847/1538-4357/abec82}}}.

\bibitem[{Tomicic} \em{et~al.}(2021){Tomicic}, {Vulcani}, {Poggianti}, {Werle},
  {Muller}, {Mingozzi}, {Gullieuszik}, {Wolter}, {Radovich}, {Moretti},
  {Franchetto}, {Bellhouse}, and {Fritz}]{Tomicic}
{Tomicic}, N.; {Vulcani}, B.; {Poggianti}, B.M.; {Werle}, A.; {Muller}, A.;
  {Mingozzi}, M.; {Gullieuszik}, M.; {Wolter}, A.; {Radovich}, M.; {Moretti},
  A.; {Franchetto}, A.; {Bellhouse}, C.; {Fritz}, J.
\newblock {GASP XXXV: Characteristics of the diffuse ionised gas in
  gas-stripped galaxies}.
\newblock {\em arXiv e-prints} {\bf 2021}, p. arXiv:2108.12433,
  \href{http://xxx.lanl.gov/abs/2108.12433}{{\normalfont
  [arXiv:astro-ph.GA/2108.12433]}}.

\bibitem[{Moretti} \em{et~al.}(2020){Moretti}, {Paladino}, {Poggianti},
  {Serra}, {Roediger}, {Gullieuszik}, {Tomi{\v{c}}i{\'c}}, {Radovich},
  {Vulcani}, {Jaff{\'e}}, {Fritz}, {Bettoni}, {Ramatsoku}, and
  {Wolter}]{Moretti2019}
{Moretti}, A.; {Paladino}, R.; {Poggianti}, B.M.; {Serra}, P.; {Roediger}, E.;
  {Gullieuszik}, M.; {Tomi{\v{c}}i{\'c}}, N.; {Radovich}, M.; {Vulcani}, B.;
  {Jaff{\'e}}, Y.L.; {Fritz}, J.; {Bettoni}, D.; {Ramatsoku}, M.; {Wolter}, A.
\newblock {GASP. XXII. The Molecular Gas Content of the JW100 Jellyfish Galaxy
  at z {\ensuremath{\sim}} 0.05: Does Ram Pressure Promote Molecular Gas
  Formation?}
\newblock {\em Astrophysical Journal} {\bf 2020}, {\em 889},~9,
  \href{http://xxx.lanl.gov/abs/1912.06565}{{\normalfont
  [arXiv:astro-ph.GA/1912.06565]}}.
\newblock
  doi:{\changeurlcolor{black}\href{https://doi.org/10.3847/1538-4357/ab616a}{\detokenize{10.3847/1538-4357/ab616a}}}.

\bibitem[Sun \em{et~al.}(2021)Sun, Ge, Luo, Yagi, Jáchym, Boselli, Fossati,
  Nulsen, Yoshida, and Gavazzi]{Sun2021}
Sun, M.; Ge, C.; Luo, R.; Yagi, M.; Jáchym, P.; Boselli, A.; Fossati, M.;
  Nulsen, P.E.J.; Yoshida, M.; Gavazzi, G.
\newblock Tales of tails: H$\alpha$--X-ray correlation,  2021,
  \href{http://xxx.lanl.gov/abs/2103.09205}{{\normalfont
  [arXiv:astro-ph.GA/2103.09205]}}.

\bibitem[{Jaff{\'e}} \em{et~al.}(2018){Jaff{\'e}}, {Poggianti}, {Moretti},
  {Gullieuszik}, {Smith}, {Vulcani}, {Fasano}, {Fritz}, {Tonnesen}, {Bettoni},
  {Hau}, {Biviano}, {Bellhouse}, and {McGee}]{Jaffe2018}
{Jaff{\'e}}, Y.L.; {Poggianti}, B.M.; {Moretti}, A.; {Gullieuszik}, M.;
  {Smith}, R.; {Vulcani}, B.; {Fasano}, G.; {Fritz}, J.; {Tonnesen}, S.;
  {Bettoni}, D.; {Hau}, G.; {Biviano}, A.; {Bellhouse}, C.; {McGee}, S.
\newblock {GASP. IX. Jellyfish galaxies in phase-space: an orbital study of
  intense ram-pressure stripping in clusters}.
\newblock {\em Monthly Notices of the Royal Astronomical Society} {\bf 2018},
  {\em 476},~4753--4764,
  \href{http://xxx.lanl.gov/abs/1802.07297}{{\normalfont
  [arXiv:astro-ph.GA/1802.07297]}}.
\newblock
  doi:{\changeurlcolor{black}\href{https://doi.org/10.1093/mnras/sty500}{\detokenize{10.1093/mnras/sty500}}}.

\bibitem[{Ignesti} \em{et~al.}(2018){Ignesti}, {Gitti}, {Brunetti},
  {O'Sullivan}, {Sarazin}, and {Wong}]{Ignesti_2018}
{Ignesti}, A.; {Gitti}, M.; {Brunetti}, G.; {O'Sullivan}, E.; {Sarazin}, C.;
  {Wong}, K.
\newblock {The mystery of the ``Kite'' radio source in Abell 2626: Insights
  from new Chandra observations}.
\newblock {\em Astronomy and Astrophysics} {\bf 2018}, {\em 610},~A89,
  \href{http://xxx.lanl.gov/abs/1712.07884}{{\normalfont
  [arXiv:astro-ph.GA/1712.07884]}}.
\newblock
  doi:{\changeurlcolor{black}\href{https://doi.org/10.1051/0004-6361/201731380}{\detokenize{10.1051/0004-6361/201731380}}}.

\bibitem[{Strickland} \em{et~al.}(2004){Strickland}, {Heckman}, {Colbert},
  {Hoopes}, and {Weaver}]{Strickland_2004a}
{Strickland}, D.K.; {Heckman}, T.M.; {Colbert}, E.J.M.; {Hoopes}, C.G.;
  {Weaver}, K.A.
\newblock {A High Spatial Resolution X-Ray and H{\ensuremath{\alpha}} Study of
  Hot Gas in the Halos of Star-forming Disk Galaxies. I. Spatial and Spectral
  Properties of the Diffuse X-Ray Emission}.
\newblock {\em Astrophysical Journal} {\bf 2004}, {\em 151},~193--236,
  \href{http://xxx.lanl.gov/abs/astro-ph/0306592}{{\normalfont
  [arXiv:astro-ph/astro-ph/0306592]}}.
\newblock
  doi:{\changeurlcolor{black}\href{https://doi.org/10.1086/382214}{\detokenize{10.1086/382214}}}.

\bibitem[{M{\"u}ller} \em{et~al.}(2021){M{\"u}ller}, {Pfrommer}, {Ignesti},
  {Moretti}, {Louren{\c{c}}o}, {Paladino}, {Jaff{\'e}}, {Gitti}, {Venturi},
  {Gullieuszik}, {Poggianti}, {Vulcani}, {Biviano}, {Adebahr}, and
  {Dettmar}]{Mueller21}
{M{\"u}ller}, A.; {Pfrommer}, C.; {Ignesti}, A.; {Moretti}, A.;
  {Louren{\c{c}}o}, A.; {Paladino}, R.; {Jaff{\'e}}, Y.; {Gitti}, M.;
  {Venturi}, T.; {Gullieuszik}, M.; {Poggianti}, B.; {Vulcani}, B.; {Biviano},
  A.; {Adebahr}, B.; {Dettmar}, R.J.
\newblock {Two striking head-tail galaxies in the galaxy cluster IIZW108:
  insights into transition to turbulence, magnetic fields, and particle
  re-acceleration}.
\newblock {\em Monthly Notices of the Royal Astronomical Society} {\bf 2021},
  {\em 508},~5326--5344,
  \href{http://xxx.lanl.gov/abs/2110.03705}{{\normalfont
  [arXiv:astro-ph.GA/2110.03705]}}.
\newblock
  doi:{\changeurlcolor{black}\href{https://doi.org/10.1093/mnras/stab2928}{\detokenize{10.1093/mnras/stab2928}}}.

\bibitem[{Gitti}(2013)]{Gitti_2013}
{Gitti}, M.
\newblock {The puzzling radio source in the cool core cluster A2626.}
\newblock {\em Monthly Notices of the Royal Astronomical Society} {\bf 2013},
  {\em 436},~L84--L88,  \href{http://xxx.lanl.gov/abs/1308.5825}{{\normalfont
  [arXiv:astro-ph.CO/1308.5825]}}.
\newblock
  doi:{\changeurlcolor{black}\href{https://doi.org/10.1093/mnrasl/slt118}{\detokenize{10.1093/mnrasl/slt118}}}.

\bibitem[{Ignesti} \em{et~al.}(2017){Ignesti}, {Gitti}, {Brunetti}, {Feretti},
  and {Giovannini}]{Ignesti_2017}
{Ignesti}, A.; {Gitti}, M.; {Brunetti}, G.; {Feretti}, L.; {Giovannini}, G.
\newblock {New JVLA observations at 3 GHz and 5.5 GHz of the ``Kite'' radio
  source in Abell 2626}.
\newblock {\em Astronomy and Astrophysics} {\bf 2017}, {\em 604},~A21,
  \href{http://xxx.lanl.gov/abs/1705.01787}{{\normalfont
  [arXiv:astro-ph.GA/1705.01787]}}.
\newblock
  doi:{\changeurlcolor{black}\href{https://doi.org/10.1051/0004-6361/201730964}{\detokenize{10.1051/0004-6361/201730964}}}.

\bibitem[{Ignesti} \em{et~al.}(2020){Ignesti}, {Shimwell}, {Brunetti}, {Gitti},
  {Intema}, {van Weeren}, {Hardcastle}, {Clarke}, {Botteon}, {Di Gennaro},
  {Br{\"u}ggen}, {Browne}, {Mandal}, {R{\"o}ttgering}, {Cuciti}, {de Gasperin},
  {Cassano}, and {Scaife}]{Ignesti_2020}
{Ignesti}, A.; {Shimwell}, T.; {Brunetti}, G.; {Gitti}, M.; {Intema}, H.; {van
  Weeren}, R.J.; {Hardcastle}, M.J.; {Clarke}, A.O.; {Botteon}, A.; {Di
  Gennaro}, G.; {Br{\"u}ggen}, M.; {Browne}, I.W.A.; {Mandal}, S.;
  {R{\"o}ttgering}, H.J.A.; {Cuciti}, V.; {de Gasperin}, F.; {Cassano}, R.;
  {Scaife}, A.M.M.
\newblock {The great Kite in the sky: A LOFAR observation of the radio source
  in Abell 2626}.
\newblock {\em Astronomy and Astrophysics} {\bf 2020}, {\em 643},~A172,
  \href{http://xxx.lanl.gov/abs/2009.11210}{{\normalfont
  [arXiv:astro-ph.GA/2009.11210]}}.
\newblock
  doi:{\changeurlcolor{black}\href{https://doi.org/10.1051/0004-6361/202039056}{\detokenize{10.1051/0004-6361/202039056}}}.

\bibitem[{Tonnesen} and {Stone}(2014)]{Tonnesen}
{Tonnesen}, S.; {Stone}, J.
\newblock {The Ties that Bind? Galactic Magnetic Fields and Ram Pressure
  Stripping}.
\newblock {\em Astrophys. J.} {\bf 2014}, {\em 795},~148.
\newblock
  doi:{\changeurlcolor{black}\href{https://doi.org/10.1088/0004-637X/795/2/148}{\detokenize{10.1088/0004-637X/795/2/148}}}.

\bibitem[{Ramos-Mart{\'i}nez} \em{et~al.}(2018){Ramos-Mart{\'i}nez},
  {G{\'o}mez}, and {P{\'e}rez-Villegas}]{Martinez}
{Ramos-Mart{\'i}nez}, M.; {G{\'o}mez}, G.C.; {P{\'e}rez-Villegas}, {\'A}.
\newblock {MHD simulations of ram pressure stripping of a disc galaxy}.
\newblock {\em Mon. Not. R. Astron. Soc.} {\bf 2018}, {\em 476},~3781--3792.
\newblock
  doi:{\changeurlcolor{black}\href{https://doi.org/10.1093/mnras/sty393}{\detokenize{10.1093/mnras/sty393}}}.

\bibitem[{Vollmer} \em{et~al.}(2007){Vollmer}, {Soida}, {Beck}, {Urbanik},
  {Chy{\.z}y}, {Otmianowska-Mazur}, {Kenney}, and {van Gorkom}]{Vollmer07}
{Vollmer}, B.; {Soida}, M.; {Beck}, R.; {Urbanik}, M.; {Chy{\.z}y}, K.T.;
  {Otmianowska-Mazur}, K.; {Kenney}, J.D.P.; {van Gorkom}, J.H.
\newblock {The characteristic polarized radio continuum distribution of cluster
  spiral galaxies}.
\newblock {\em Astron. Astrophys.} {\bf 2007}, {\em 464},~L37--L40.
\newblock
  doi:{\changeurlcolor{black}\href{https://doi.org/10.1051/0004-6361:20066980}{\detokenize{10.1051/0004-6361:20066980}}}.

\bibitem[Dursi and Pfrommer(2008)]{Dursi}
Dursi, L.J.; Pfrommer, C.
\newblock Draping of Cluster Magnetic Fields over Bullets and
  Bubbles{\textemdash}Morphology and Dynamic Effects.
\newblock {\em Astrophys. J.} {\bf 2008}, {\em 677},~993--1018.
\newblock
  doi:{\changeurlcolor{black}\href{https://doi.org/10.1086/529371}{\detokenize{10.1086/529371}}}.

\bibitem[{Pfrommer} and {Dursi}(2010)]{Pfrommer10}
{Pfrommer}, C.; {Dursi}, L.J.
\newblock {Detecting the orientation of magnetic fields in galaxy clusters}.
\newblock {\em Nature Physics} {\bf 2010}, {\em 6},~520--526.
\newblock
  doi:{\changeurlcolor{black}\href{https://doi.org/10.1038/nphys1657}{\detokenize{10.1038/nphys1657}}}.

\bibitem[{Ruszkowski} \em{et~al.}(2014){Ruszkowski}, {Br{\"u}ggen}, {Lee}, and
  {Shin}]{Ruszkowski}
{Ruszkowski}, M.; {Br{\"u}ggen}, M.; {Lee}, D.; {Shin}, M.S.
\newblock {Impact of Magnetic Fields on Ram Pressure Stripping in Disk
  Galaxies}.
\newblock {\em Astrophys. J.} {\bf 2014}, {\em 784},~75.
\newblock
  doi:{\changeurlcolor{black}\href{https://doi.org/10.1088/0004-637X/784/1/75}{\detokenize{10.1088/0004-637X/784/1/75}}}.

\bibitem[{Berlok} and {Pfrommer}(2019)]{Berlok}
{Berlok}, T.; {Pfrommer}, C.
\newblock {The impact of magnetic fields on cold streams feeding galaxies}.
\newblock {\em Mon. Not. R. Astron. Soc.} {\bf 2019}, {\em 489},~3368--3384.
\newblock
  doi:{\changeurlcolor{black}\href{https://doi.org/10.1093/mnras/stz2347}{\detokenize{10.1093/mnras/stz2347}}}.

\bibitem[Springel(2010)]{Springel}
Springel, V.
\newblock Moving-mesh hydrodynamics with the AREPO code.
\newblock {\em Proceedings of the International Astronomical Union} {\bf 2010},
  {\em 6},~203–206.
\newblock
  doi:{\changeurlcolor{black}\href{https://doi.org/10.1017/S1743921311000378}{\detokenize{10.1017/S1743921311000378}}}.

\bibitem[{Pakmor} and {Springel}(2013)]{Pakmor13}
{Pakmor}, R.; {Springel}, V.
\newblock {Simulations of magnetic fields in isolated disc galaxies}.
\newblock {\em Mon. Not. R. Astron. Soc.} {\bf 2013}, {\em 432},~176--193.
\newblock
  doi:{\changeurlcolor{black}\href{https://doi.org/10.1093/mnras/stt428}{\detokenize{10.1093/mnras/stt428}}}.

\bibitem[Pakmor \em{et~al.}(2015)Pakmor, Springel, Bauer, Mocz, Munoz, Ohlmann,
  Schaal, and Zhu]{Pakmor15}
Pakmor, R.; Springel, V.; Bauer, A.; Mocz, P.; Munoz, D.J.; Ohlmann, S.T.;
  Schaal, K.; Zhu, C.
\newblock {Improving the convergence properties of the moving-mesh code AREPO}.
\newblock {\em Mon. Not. R. Astron. Soc.} {\bf 2015}, {\em 455},~1134--1143.
\newblock
  doi:{\changeurlcolor{black}\href{https://doi.org/10.1093/mnras/stv2380}{\detokenize{10.1093/mnras/stv2380}}}.

\bibitem[{Sparre} \em{et~al.}(2020){Sparre}, {Pfrommer}, and {Ehlert}]{Sparre}
{Sparre}, M.; {Pfrommer}, C.; {Ehlert}, K.
\newblock {Interaction of a cold cloud with a hot wind: the regimes of cloud
  growth and destruction and the impact of magnetic fields}.
\newblock {\em arXiv e-prints} {\bf 2020}.

\bibitem[{Gronke} and {Oh}(2018)]{Gronke}
{Gronke}, M.; {Oh}, S.P.
\newblock {The growth and entrainment of cold gas in a hot wind}.
\newblock {\em Mon. Not. R. Astron. Soc.} {\bf 2018}, {\em 480},~L111--L115.
\newblock
  doi:{\changeurlcolor{black}\href{https://doi.org/10.1093/mnrasl/sly131}{\detokenize{10.1093/mnrasl/sly131}}}.

\bibitem[{Li} \em{et~al.}(2020){Li}, {Hopkins}, {Squire}, and
  {Hummels}]{Li2020}
{Li}, Z.; {Hopkins}, P.F.; {Squire}, J.; {Hummels}, C.
\newblock {On the survival of cool clouds in the circumgalactic medium}.
\newblock {\em Mon. Not. R. Astron. Soc.} {\bf 2020}, {\em 492},~1841--1854.
\newblock
  doi:{\changeurlcolor{black}\href{https://doi.org/10.1093/mnras/stz3567}{\detokenize{10.1093/mnras/stz3567}}}.

\bibitem[McDonald \em{et~al.}(2012)McDonald, Veilleux, and Rupke]{Donald2012}
McDonald, M.; Veilleux, S.; Rupke, D.S.N.
\newblock OPTICAL SPECTROSCOPY OF HALPHA FILAMENTS IN COOL CORE CLUSTERS:
  KINEMATICS, REDDENING, AND SOURCES OF IONIZATION.
\newblock {\em The Astrophysical Journal} {\bf 2012}, {\em 746},~153.
\newblock
  doi:{\changeurlcolor{black}\href{https://doi.org/10.1088/0004-637x/746/2/153}{\detokenize{10.1088/0004-637x/746/2/153}}}.

\bibitem[{Olivares} \em{et~al.}(2019){Olivares}, {Salome}, {Combes}, {Hamer},
  {Guillard}, {Lehnert}, {Polles}, {Beckmann}, {Dubois}, {Donahue}, {Edge},
  {Fabian}, {McNamara}, {Rose}, {Russell}, {Tremblay}, {Vantyghem}, {Canning},
  {Ferland}, {Godard}, {Peirani}, and {Pineau des Forets}]{Olivares2019}
{Olivares}, V.; {Salome}, P.; {Combes}, F.; {Hamer}, S.; {Guillard}, P.;
  {Lehnert}, M.D.; {Polles}, F.L.; {Beckmann}, R.S.; {Dubois}, Y.; {Donahue},
  M.; {Edge}, A.; {Fabian}, A.C.; {McNamara}, B.; {Rose}, T.; {Russell}, H.R.;
  {Tremblay}, G.; {Vantyghem}, A.; {Canning}, R.E.A.; {Ferland}, G.; {Godard},
  B.; {Peirani}, S.; {Pineau des Forets}, G.
\newblock {Ubiquitous cold and massive filaments in cool core clusters}.
\newblock {\em Astronomy and Astrophysics} {\bf 2019}, {\em 631},~A22,
  \href{http://xxx.lanl.gov/abs/1902.09164}{{\normalfont
  [arXiv:astro-ph.GA/1902.09164]}}.
\newblock
  doi:{\changeurlcolor{black}\href{https://doi.org/10.1051/0004-6361/201935350}{\detokenize{10.1051/0004-6361/201935350}}}.

\bibitem[{Morabito} \em{et~al.}(2021){Morabito}, {Jackson}, {Mooney},
  {Sweijen}, {Badole}, {Kukreti}, {Venkattu}, {Groeneveld}, {Kappes},
  {Bonnassieux}, {Drabent}, {Iacobelli}, {Croston}, {Best}, {Bondi},
  {Callingham}, {Conway}, {Deller}, {Hardcastle}, {McKean}, {Miley}, {Moldon},
  {R{\"o}ttgering}, {Tasse}, {Shimwell}, {van Weeren}, {Anderson}, {Asgekar},
  {Avruch}, {van Bemmel}, {Bentum}, {Bonafede}, {Brouw}, {Butcher}, {Ciardi},
  {Corstanje}, {Coolen}, {Damstra}, {de Gasperin}, {Duscha}, {Eisl{\"o}ffel},
  {Engels}, {Falcke}, {Garrett}, {Griessmeier}, {Gunst}, {van Haarlem},
  {Hoeft}, {van der Horst}, {J{\"u}tte}, {Kadler}, {Koopmans}, {Krankowski},
  {Mann}, {Nelles}, {Oonk}, {Orru}, {Paas}, {Pandey}, {Pizzo},
  {Pandey-Pommier}, {Reich}, {Rothkaehl}, {Ruiter}, {Schwarz}, {Shulevski},
  {Soida}, {Tagger}, {Vocks}, {Wijers}, {Wijnholds}, {Wucknitz}, {Zarka}, and
  {Zucca}]{Morabito_2021}
{Morabito}, L.K.; {Jackson}, N.J.; {Mooney}, S.; {Sweijen}, F.; {Badole}, S.;
  {Kukreti}, P.; {Venkattu}, D.; {Groeneveld}, C.; {Kappes}, A.; {Bonnassieux},
  E.; {Drabent}, A.; {Iacobelli}, M.; {Croston}, J.H.; {Best}, P.N.; {Bondi},
  M.; {Callingham}, J.R.; {Conway}, J.E.; {Deller}, A.T.; {Hardcastle}, M.J.;
  {McKean}, J.P.; {Miley}, G.K.; {Moldon}, J.; {R{\"o}ttgering}, H.J.A.;
  {Tasse}, C.; {Shimwell}, T.W.; {van Weeren}, R.J.; {Anderson}, J.M.;
  {Asgekar}, A.; {Avruch}, I.M.; {van Bemmel}, I.M.; {Bentum}, M.J.;
  {Bonafede}, A.; {Brouw}, W.N.; {Butcher}, H.R.; {Ciardi}, B.; {Corstanje},
  A.; {Coolen}, A.; {Damstra}, S.; {de Gasperin}, F.; {Duscha}, S.;
  {Eisl{\"o}ffel}, J.; {Engels}, D.; {Falcke}, H.; {Garrett}, M.A.;
  {Griessmeier}, J.; {Gunst}, A.W.; {van Haarlem}, M.P.; {Hoeft}, M.; {van der
  Horst}, A.J.; {J{\"u}tte}, E.; {Kadler}, M.; {Koopmans}, L.V.E.;
  {Krankowski}, A.; {Mann}, G.; {Nelles}, A.; {Oonk}, J.B.R.; {Orru}, E.;
  {Paas}, H.; {Pandey}, V.N.; {Pizzo}, R.F.; {Pandey-Pommier}, M.; {Reich}, W.;
  {Rothkaehl}, H.; {Ruiter}, M.; {Schwarz}, D.J.; {Shulevski}, A.; {Soida}, M.;
  {Tagger}, M.; {Vocks}, C.; {Wijers}, R.A.M.J.; {Wijnholds}, S.J.; {Wucknitz},
  O.; {Zarka}, P.; {Zucca}, P.
\newblock {Sub-arcsecond imaging with the International LOFAR Telescope I.
  Foundational calibration strategy and pipeline}.
\newblock {\em arXiv e-prints} {\bf 2021}, p. arXiv:2108.07283,
  \href{http://xxx.lanl.gov/abs/2108.07283}{{\normalfont
  [arXiv:astro-ph.IM/2108.07283]}}.

\end{thebibliography}

\end{document}